\documentclass[showpacs,aps,prl,twocolumn]{revtex4}

\usepackage{graphicx}
\usepackage{dcolumn}
\usepackage{bm}
\usepackage{amssymb}
\usepackage{amsfonts,amsmath}
\usepackage[dvips]{color}

\usepackage[final]{changes}
\definechangesauthor[name={Francesco Papoff}, color=red]{fp}
\definechangesauthor[name={Duncan McArthur}, color=blue]{dm}
\definechangesauthor[name={Alison Yao}, color=violet]{ay}
\newcommand{\beq}{\begin{equation}}
\newcommand{\eeq}{\end{equation}}
\newcommand{\bea}{\begin{eqnarray*}}
\newcommand{\eea}{\end{eqnarray*}}
\newcommand{\bnea}{\begin{eqnarray}}
\newcommand{\enea}{\end{eqnarray}}
\newcommand{\p}{\partial}
\newcommand{\dsp}{\displaystyle}
\newcommand{\ba}{\begin{array}{cc}}
\newcommand{\ea}{\end{array}}

\newcommand{\tr}{\mathrm{T}}

\begin{document}


\title{Scattering of light with angular momentum from 
an array of particles} 

\author{Duncan McArthur}
\author{Alison M. Yao}
\author{Francesco Papoff}
 \email{f.papoff@strath.ac.uk}
\affiliation{%
Department of Physics, SUPA, University of Strathclyde, 
107 Rottenrow, Glasgow G4 0NG, UK.
}%

\date{\today}

\begin{abstract}
Understanding the scattering properties of various media is of critical 
importance in many applications, from secure, high-bandwidth communications to 
extracting information about biological and mineral particles dissolved in sea 
water. 
In this paper we demonstrate how beams carrying orbital angular momentum (OAM) 
can be used to detect the presence of symmetric or chiral subsets of particles 
in disordered media. Using a generalized Mie theory we calculate analytical 
expressions for quasi-monochromatic structured light scattered by dilute 
distributions of micro- and nanoparticles. 
These allow us to determine the angular momentum of the scattered field as a 
function of the angular momentum of the incident beam and of the spatial 
distributions of scattering particles. Our numerical results show that we can 
distinguish structured from random distributions of particles\replaced[id=dm]{, 
even when the number density of ordered particles is a few percent of the total 
distribution}{down to a few percent}. 
\replaced[id=dm]{We also find that the signal-to-noise ratio, in the forward direction, is 
equivalent for all orders of the Laguerre-Gaussian modes in relatively dense 
(but still dilute) distributions, making them an ideal basis to encode and
transmit multiplexed signals.}{
We also show that the signal-to-noise ratio for forward scattering can be improved 
by using incident beams with higher OAM values.}
\end{abstract}

\maketitle

%
\section{\label{sec:intro} Introduction}
Light carrying orbital angular momentum (OAM) has attracted a great deal of 
interest due to its potential to enhance many applications, including 
high-resolution imaging \cite{klar00}, and optical trapping and manipulation of 
micro- and nanoparticles\cite{he95}. It is also of significant interest as a 
resource in high capacity quantum communication and information systems. Unlike 
the spin angular momentum (SAM) associated with circular polarization, OAM is 
not restricted to two orthogonal states. Large amounts of information can then 
be carried by single photons, opening the door to new protocols for quantum key 
distribution with significantly increased data transfer rates~\cite{tyler09}. 

As any communications system relies on the fidelity of the signal after 
propagation, it is essential to understand the effect of the medium that they 
travel through. In this case we focus on the effect of propagation through 
scattering media. Whether this be due to scattering by aerosols in the 
atmosphere or by phytoplankton or mineral particles in natural waters~\cite{
mobley94}, scattering will contribute both to the coherence and attenuation of 
the signal, resulting in a loss of information and a reduction in the fidelity 
of the system.

While scattering is considered to be detrimental for applications in 
information transfer, the ability to measure the power of the OAM states of the 
scattered light~\cite{Leach04,berkhout2010,Dudley2013,boyd13,Gu2018,fontaine2019} 
may provide a source of information about the properties of the scattering 
medium, from the size and type of the particulates to their geometrical 
distribution, that has not yet been exploited. 
This may be of particular benefit for environmental sensing and metrology, and 
ocean transmissometry, and may even find applications in biological imaging. 
Another area of potential application is nanophotonics, where some of the 
unique features of the scattering of light carrying OAM can reveal the fraction 
of ordered and disordered nanoparticles \replaced[id=dm]{within}{inside} complex 
nanostructures.

In this paper we present a theory for quasi-monochromatic structured light 
scattered by dilute distributions of micro- and nanoparticles that allows us to 
determine the angular momentum components of the scattered field in the far 
field zone as a function of the total angular momentum of the incident beam 
along its axis and of the dielectric properties and spatial distributions of 
the scattering particles. 
For paraxial incident fields, we show that it is possible to detect the 
presence of \deleted[id=dm]{symmetric} subsets of \added[id=dm]{symmetric and chiral}
particles \deleted[id=dm]{and chirality} simply by controlling 
the OAM of the incident beams and measuring the resultant OAM of the scattered 
fields in the far field. This may help us to detect objects hidden in 
scattering media~\cite{cochenour17} or to visualize flows resulting in 
geometrically ordered concentrations of particles~\cite{balkovsky01}. Here we 
consider both the case of particles distributed over volumes with dimensions 
several orders of magnitude larger than the wavelength of light, and the case 
of particles distributed over volumes with \replaced[id=dm]{dimensions of}{sides} 
a few ten\deleted[id=dm]{th}s of the wavelength. 
The first case is typical of marine optics~\cite{mobley94} and atmospheric 
science, or of nanophotonic experiments with nanoparticles in solutions.  The 
second case is typical of experiments involving particles held in optical traps 
or forming artificial nanophotonics structures~\cite{Grier2003,Polin2006,
Leonardo07,Curran2010}.  We show that 
controlling the OAM of the incident beam and measuring the power of the light 
scattered in different OAM states allows us to identify the presence of subsets 
of particles forming polygonal or chiral structures in both cases.
We also demonstrate that OAM can be used to improve the signal to noise ratio, 
a result that can be useful in reducing the error due to scattering both in 
measures of absorption, and in communications applications.

As the theory is necessarily very mathematical, we start by giving an outline 
of our method and our main theoretical results. We then present a number of 
numerical results highlighting the main features and capabilities of our theory.
For the sake of simplicity and to facilitate the comparisons between theory and 
experiments, our numerical results have been obtained using distributions of 
gold nanospheres as these are used in many experiments. However, we stress that 
our theory is applicable to \emph{any type of scattering particles and host medium}, 
provided that multiple scattering can be neglected. For the interested reader, 
we then present a detailed outline of our theory.

\section{\label{sec:over} Overview of Theory and Main Results}
We consider an experimental set-up, as shown in Fig.~\ref{fig:setup}, where 
light beams propagating along the $z$-axis are incident on distributions of 
micro- and nanoparticles with dielectric permitivitty $\varepsilon_r$ and 
magnetic permeability $\mu_r$ immersed in a uniform isotropic dielectric medium 
with different \replaced[id=dm]{electromagnetic properties}{$\varepsilon_r, \mu_r$}. 
Using a generalized Mie theory, we calculate the far field scattering of both 
random and symmetric distributions of particles in which multiple scattering can 
be neglected and we can safely consider the total scattered field as the 
coherent sum of the fields scattered by the individual particles. 
Our theory is applicable to any type of scattering particle and host medium and 
is developed for arbitrary particles and beams.

From a physical point of view, the key issue is that we expand the field 
scattered by particles in different positions in electric and magnetic 
multipolar waves~\cite{biedenharn85}, $\mathcal{S}_{jm}^{H }$ and 
$\mathcal{S}_{jm}^{E }$, where, for each particle, \deleted[id=dm]{$j(j+1)$ 
with} $j=1,2,\dots$ \replaced[id=dm]{is}{and $m=  \pm 1, \pm 2, \dots$ are} the 
total angular momentum with respect to the center of the particle and 
\added[id=dm]{$m=\pm 1,\pm 2,\dots$ is} its component parallel to the direction 
of propagation of the incident beam in non-dimensional units. 
However $j_z$, the total angular momentum of the scattered waves 
$\mathcal{S}_{jm}^{H }$  and $\mathcal{S}_{jm}^{E }$  {\em along the axis of 
the incident beam}, is not $m$ and depends on the distance between the centers 
of the off-axis scattering particles and the axis of the incident beam. 
The essential features of scattering from dilute\deleted[id=dm]{d} distributions of particles 
depend on this property and on the fact that waves scattered by different 
particles are coherent. \replaced[id=dm]{Due to}{Because of} the coherent addition of the scattered waves, 
structured incident beams can be used to induce collective scattering that 
reveals the spatial structure of the particles' distributions. More 
specifically, one can detect the presence of subsets of particles arranged with 
the roto-reflection symmetry of regular polygons with $N$ sides, or \deleted[id=dm]{on }chiral 
structures. It is also possible to detect the position of the \added[id=dm]{symmetry}
axes of two and three dimensional arrays and chiral structures by displacing the 
\added[id=dm]{particles with respect to the} incident beam's axis and measuring 
the spread of the scattered light power over 
the component of OAM along the beam axis of the scattered field, $l_z$. The 
resolution on the transverse position of symmetry axes, or the centers of 
chiral structures, is of a few percent of the transverse dimension of the 
incident beam.

For the incident beams considered here, the main theoretical results depend on 
the angular momentum of the incident beam and the types and distributions of 
the particles. 
The main results \replaced[id=dm]{can be summarized as follows}{are}:
\begin{itemize}
\item
{\bf 
The multipolar waves 
$\bm{\mathcal{S}_{jm}^{H}}$ and $\bm{\mathcal{S}_{jm}^{E}}$, scattered by every particle, 
combine in a scattered wave with total angular momentum along the axis of the beam 
$\bm{j_z = u+\ell+m}$.}
In non-dimensional units, $j_z=0, \pm 1, \pm 2 \dots$ is the total angular 
momentum along the beam's axis $z$; $\ell=0,\pm 1, \pm 2,\dots$ is the 
component of OAM of the incident Laguerre-Gauss beam along its axis; $u$ 
defines the spatial harmonics of the multipole-multipole distributions, 
$\exp{(iu\varphi)}$, where $\varphi$ is the azimuthal angle.
\item
{\bf For light, the component of the spin along the propagation axis takes 
three values:
$\bm{s_z=0}$ for polarization $\mathbf{\hat{z}}$, and $\bm{s_z = \pm 1}$ for 
polarization $\mathbf{\hat{e}}_{\pm}$. The component of OAM along the beam axis 
of the scattered field, $\bm{l_z} $, can also take three values for any given 
$\bm{j_z}$, as $\bm{l_z = j_z-s_z}$}.
\item
{\bf In the forward and backward directions, only the terms with both 
$\bm{u=-\ell}$ and $\bm{s_z=m=\pm 1}$ do not vanish.}
\item
{\bf In the forward direction most particles scatter in phase.} \replaced[id=dm]{T}{since t}he 
positions of the particles only affect the scattering amplitude in the forward 
direction through the slowly varying amplitude of the incident field.
\item
{\bf In the backward direction the scattered fields of most particles will 
cancel out.} \replaced[id=dm]{This is due to}{since there are} rapidly varying 
phase terms that depend on the coordinates of the particles along the direction 
of propagation of the incident beam. The exceptions to this are particles that 
either all lie on a plane orthogonal to the propagation axis or are 
periodically distributed along the direction of propagation of the incident 
beam, as the fields scattered by these particles add in phase in the backward 
direction.
\end{itemize}

\section{\label{sec:results} Numerical results}
\begin{figure}[ht!]
\includegraphics[clip=false]{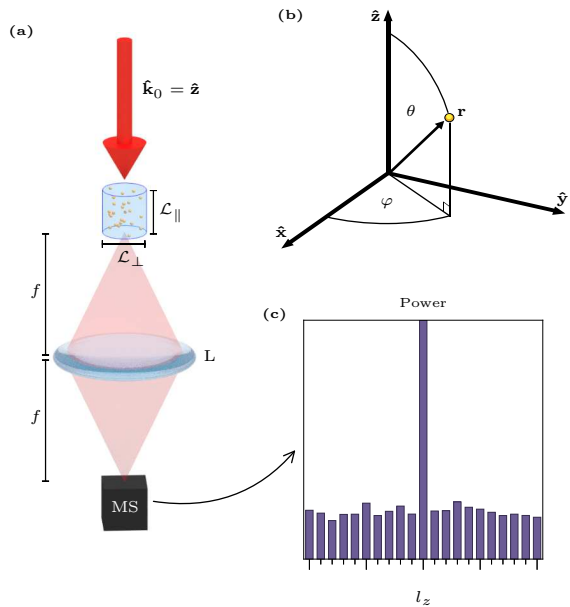}
  \caption{ \label{fig:setup} Schematic of a basic experimental setup (a)
  where an incident beam propagating along the direction $\mathbf{\hat{k}}_0$
  interacts with a distribution of particles within a cylindrical volume 
  defined by the dimensions $\mathcal{L}_{\perp}$ and $\mathcal{L}_{\parallel}$.
  Light scattered by the particles is collected by a lens (L), with focal length
  $f$, and coupled to an orbital angular momentum (OAM) mode sorter (MS). The 
  resulting signals are then processed to determine the OAM spectrum ($l_z$) of the 
  scattered light (c). The coordinate system we adopt is shown in (b) for a 
  particle located at the position $\mathbf{r}$.
  }
\end{figure}

In this section we show how incident beams carrying OAM can be used to extract 
information about the spatial structure of distributions of scattering 
particles.
A schematic \replaced[id=dm]{showing the essential components of the type of 
experimental setup we envisage}
{of our proposed experimental set-up} is shown in Fig.~\ref{fig:setup}: 
with respect to the theory developed here, the key features are provided by the 
OAM sorters that determine the OAM of the incident beam and 
\replaced[id=dm]{enable measurements of}{allow one to measure} the distribution 
of the scattered field power over different OAM values~\cite{Leach04,berkhout2010,
Dudley2013,boyd13,Gu2018,fontaine2019}.
\replaced[id=dm]{T}{Note that t}his problem has several very different spatial scales that need to be 
considered carefully: the wavelength $\lambda$ of the incident and scattered 
fields, the largest dimension of the particles, $R_t$, the transverse and 
longitudinal dimensions of the distribution, $\mathcal{L}_\perp$ and 
$\mathcal{L}_\parallel$, as well as the transverse dimension of the incident 
beam, $w_0$, and its coherence length $\mathcal{L}_c$. 
We consider only arrays of particles with inter-particle distances such that 
multiple scattering can be neglected~\cite{berg08a} and with 
$\mathcal{L}_\parallel < \mathcal{L}_c$, where the total scattered field is the 
coherent sum of all the fields scattered by each particle. 
We note that the latter condition is not a limit to this theory. For the case 
$\mathcal{L}_\parallel > \mathcal{L}_c$ the distribution can be divided into 
sections of length $\mathcal{L}_c$ with the total field being the 
incoherent sum of the resulting fields from each section.

Our numerical results are obtained by considering paraxial Laguerre-Gaussian 
beams and spheres with radii $r \ll w_0, z_R$, where $w_0$ is the beam waist 
and $z_R$ is the corresponding Rayleigh parameter, $z_R=k w_0^2/2$.
As we show later in the theory section, only corrections to the paraxial fields 
of order $k^{-2}$ affect the scattered fields; for the cases considered here, 
corrections of this order can be safely neglected. 
For simplicity, we consider only the ``purely azimuthal'' $LG_{\ell p}$ modes 
($p=0$),\deleted[id=dm]{ with linear polarization,} although the radial 
modes could provide additional information, as discussed in Sec.~\ref{sec:theory}. 

As mentioned above, our theory is applicable to \emph{any} type of scattering 
particles and host medium. 
\replaced[id=dm]{However, again for the sake of simplicity,}{For the sake of 
simplicity, however,} and to facilitate the comparisons between theory and 
experiments, our numerical results have been obtained using distributions of 
gold nanospheres with radii of $80$~nm. We use the Lorentz-Drude dielectric 
function~\cite{Rakic98} to model the electromagnetic response of the particles. 
As a result of their size and dielectric function these particles have a 
spectrally resolved Mie-type resonance with a peak around $670$~nm for the 
scattered fields. Using a different dielectric function would affect 
the position of the resonance~\cite{imura14a}, but would not affect the ability 
to identify spatial properties of the particles' distributions which depends 
on the general properties of the scattering process.

\subsection{\label{sec:random} Ordered and disordered particle distributions}
We first consider particles distributed over volumes with 
\replaced[id=dm]{dimensions}{sides} much larger than the wavelength, as is the 
case in nanophotonics with large numbers of nanoparticles in \replaced[id=dm]
{colloids}{solutions} or in marine optics. In particular, we investigate: 
\begin{enumerate}
\item Nanoparticles randomly distributed within a cylindrical volume, with 
  radius of cross-section $\propto w_0$; 
\item Particles regularly distributed, on layers orthogonal to the $z$-axis, 
  along lines with the symmetry of regular $n$-sided polygons;
\item Combinations of the previous two types of distributions.
\end{enumerate}
Information on the distribution of particles is most easily extracted from 
either backward or forward scattering, so we consider these simple cases first. 
We use the index $t$ to label the spheres and the azimuthal moments of the 
particles' distribution, defined as $|\sum_t \exp{(i u \varphi_t)}|/N$, to 
identify random, partially ordered and ordered distributions. 
The azimuthal moments are especially useful when considering two dimensional 
structures and for forward scattering, where the scattered fields depends very 
weakly on the position of the particles along the $z$ (beam) axis.
For identical particles, the azimuthal moments are proportional to the volume 
integral of the azimuthal Fourier components of the particles' distributions, 
as we show in Sec.~\ref{sec:theory}.
%
\begin{figure*}[ht!]
\includegraphics[clip=false]{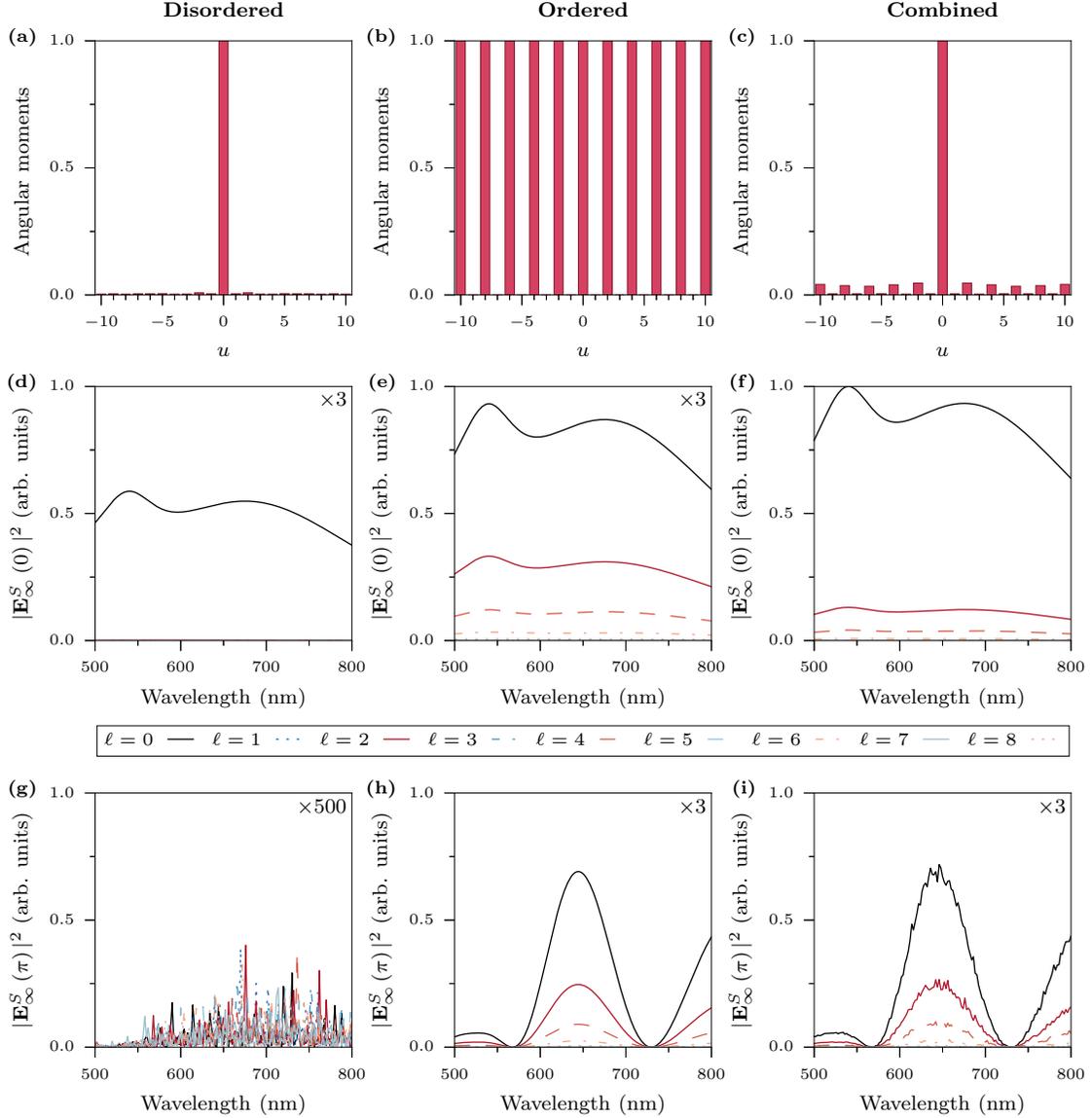}
  \caption{ \label{fig_Scat} From left to right: $N=4\times 10^4$ particles randomly
  distributed; $1.6\times 10^3$ particles distributed with $2$-fold symmetry, 
  over two parallel lines; $4.16\times 10^4$ particles obtained by combining the
  previous two distributions. (a-c): plots of the azimuthal  moments of the 
  particles' distribution, $|\sum_t \exp{(i u \varphi_t)}|/N$, for 
  the three distributions of gold nanoparticles with $80$nm radius; for random distributions, the 
  $u=0$ moment is always much larger than the others; for particles distributed with an even
  ($2$-fold) symmetry the odd azimuthal moments vanish; the mixed distribution shows a dominant $0$
  moment with weak even moments. Forward (d-f) and backward (g-i) scattering for the three 
  distributions above for incident beams with OAM ($\ell$) ranging from $0$ to $8$. Note that 
  for identical particles, forward and backward scattering only happen when $u=-\ell$.
  }
\end{figure*}
\replaced[id=dm]{Along}{In} the top row of Fig.~\ref{fig_Scat} we plot the 
azimuthal moments of the distribution of particles: for the random distribution 
the $0^\mathrm{th}$ order moment is dominant, while for the ordered 
distribution with $2$-fold symmetry, the even order moments are all of 
\replaced[id=dm]{equal}{the same} amplitude. The \replaced[id=dm]{presence}
{effect} of the $2$-fold symmetry is clearly evident even when 
the \added[id=dm]{number of} ordered particles \replaced[id=dm]{is}
{are} only $4\%$ of the \replaced[id=dm]{total distribution}{randomly distributed particles}.
In the second and third rows of Fig.~\ref{fig_Scat} we plot the forward and 
backward scattering, $\theta=0,\pi$ respectively,
for incident beams with OAM ($\ell$) ranging from $0$ to $8$ for each of these 
distributions.
Our results illustrate that only the terms with $u=-\ell$ and either 
$\mathbf{\hat{e}}_+,m= 1$, or $\mathbf{\hat{e}}_-, m= -1$ do not vanish.
Note that the forward scattering intensity shown is proportional to the modulus 
squared of the sum of the scattering amplitudes of the $jm$ electric and 
magnetic multipoles in the forward direction due to the $-\ell$ spatial 
harmonics of the distribution.
For forward scattering, the scattered fields only depend on the positions of 
the particles over the long length scales $w_0$ and $z_R$ of the paraxial 
incident beam, so many particles add up in phase. 
For backward scattering, however, the scattered fields depend on the 
$z$-coordinates of the particles along the beam's axis over the short length 
scale $\lambda/2$. 
The relative phases depend on the wavelength of the incident field and for most 
wavelengths the backward scattered fields of random distributions are not in 
phase. There are, however, some\deleted[id=dm]{ values of the} wavelengths for 
which partial coherence gives rise to larger values of the backward scattered 
field. 
Fields scattered by particles lying on a plane orthogonal to the beam axis 
are in phase, hence we observe interference fringes for symmetric 
distributions over two or more planes orthogonal to the beam axis. 
As the fields scattered by a random distribution of particles \added[id=dm]{effectively} 
cancel out at most wavelengths, these interference fringes 
\replaced[id=dm]{dominate the scattering response even when the symmetric 
distribution comprises a small fraction of the total number of particles.}{are 
still clearly distinguishable even when immersed in 
much larger numbers of randomly distributed particles}. 
\replaced[id=dm]{Therefore, by scanning the wavelength and measuring the 
interference fringes in the back-scattered field the presence of ordered 
particles, on planes orthogonal to, and periodically spaced along the incident 
beam axis, can be readily identified.}
{Measuring the interference fringes in the back-scattered field by scanning the 
wavelength then allows one to detect the presence in the distribution of 
particles of sets of planes that are periodically stacked and orthogonal to the 
incident beam's axis.}

In some experiments, weak scattered fields make the use of detection angles 
larger than those necessary to estimate forward and backward scattering. 
In order to draw comparison with \added[id=dm]{such} experiments we consider 
now measurements made \added[id=dm]{collecting light scattered} within a cone
of half-angle $5^\circ$ in both the forward and backward directions. 
Experimentally, these calculations would correspond to measures made using OAM 
mode sorters~\cite{Leach04,berkhout2010,Dudley2013,boyd13,Gu2018,fontaine2019}.

\begin{figure*}[ht!]
\includegraphics[clip=false]{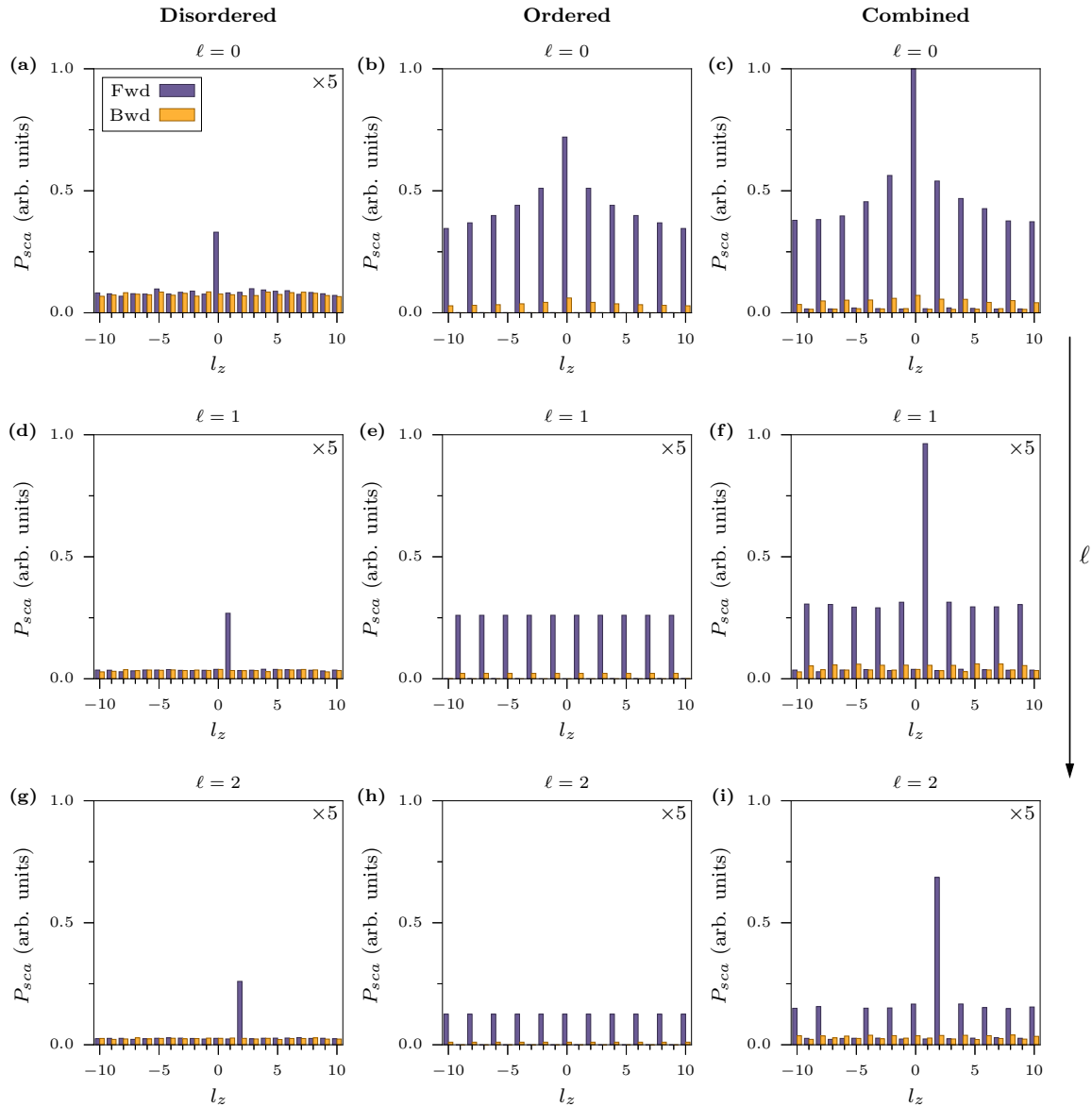}
  \caption{\label{fig_OAM} Power of scattered light with defined OAM 
  within a small cone, with a half-angle of $5^\circ$, in both the forward 
  and backward directions.
  Each column corresponds to the same distributions as in Fig~\ref{fig_Scat}.
  The rows show results for different incident beams ($LG_{\ell p}$) 
  with $p=0$, $\ell =0,1,2$. In each figure the purple (yellow) bars represent the 
  forward (backward) scattered power.
  }
\end{figure*}
In Fig.~\ref{fig_OAM} we plot the \replaced[id=dm]{scattered power}{intensity 
of light} with specific values of the OAM $l_z$ emitted in both the forward and 
backward directions.
As a consequence of the theoretical properties given above, we find that 
scattered waves with an arbitrary value of $l_z$ are originated by azimuthal 
harmonics of the multipole-multipole distributions with $u=l_z+s_z-(\ell +m)$. 
In the local plane wave approximation $m= \pm 1$ when the incident beam has 
\added[id=dm]{circular} polarization $\mathbf{\hat{e}}_\pm$.
For wide angles, for any given pair of values of $\ell$ and $l_z$, the
azimuthal harmonics that contribute to the scattered power for incident beam  
polarization $\mathbf{\hat{e}}_\mp$ are $u=l_z-\ell,l_z-\ell\pm 1,l_z-\ell\pm 2$, 
while all these five azimuthal harmonics contribute to the scattered power for 
incident beam linear polarization. For narrow detection cones in the forward or 
backward direction, we see from Eqs.~(\ref{eq:small_theta_m}-\ref{eq:small_theta_n}) 
that the dominant terms are those with $u=l_z-\ell$ and 
the same polarization as the incident beam. 
This explains why the dominant peak, corresponding to largest harmonic with 
$u=0$, is always observed at $l_z = \ell$.

\added[id=dm]{We also remark that higher order incident beams have an 
equivalent signal-to-noise ratio, compared to a fundamental Gaussian mode, for 
suitably dense (but still dilute) distributions of particles. This is to be 
expected, as the scattering process is consistent for all values of the OAM 
of the incident beam. However, further investigation would be required for 
particles with dimensions greater than, or comparable to, the wavelength of the 
incident light.}

\subsection{\label{sec:array} Regular arrays of particles}
We now investigate the application of OAM beams to nanophotonic arrays of a few 
wavelengths in size. While the relation between azimuthal Fourier components of 
the partcles' distribution and $l_z$ is simpler for narrow detection cones, as 
discussed above, for arrays of this size this approach requires the use of very 
tightly focused beams in order to maintain a good overlap between the incident 
beams and the structure for all values of $\ell$, and detect the contribution 
of the azimuthal moments with higher $u$. Theoretically, this can be 
investigated using the general theory developed in this paper together with the 
beam expansion coefficients of non-paraxial beams~\cite{gutierrez-cuevas18a}. 
Alternatively, we can use a Gaussian beam, with $\ell=0$, which always has a 
good overlap with small structures, and increase the detection angle. 
In this way the presence of azimuthal order \added[id=dm]{in the distribution} 
is reflected in the OAM states of the scattered field. 
Measuring the distribution of scattered power on the OAM states can be realized 
by placing the particles and the sorter in the focal planes of a lens, so that 
the sorter can separate the OAM states as in paraxial regime~\cite{berkhout2010}.

\begin{figure}[ht!]
\includegraphics[clip=false]{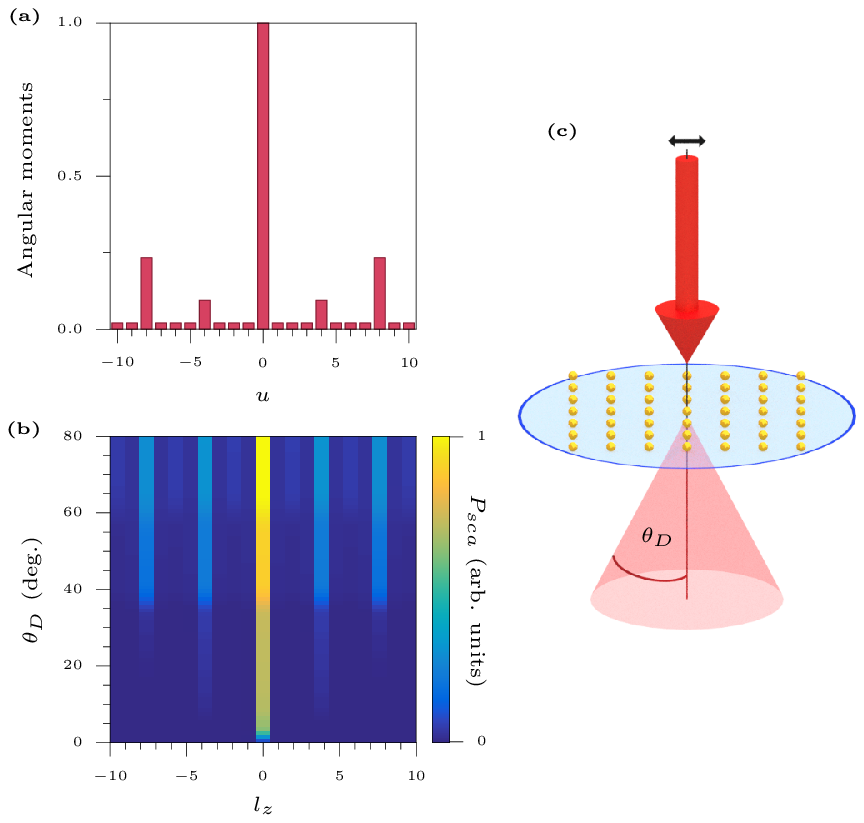}
  \caption{ \label{fig:array_angle} Effect of increasing the angle of detection,
  $\theta_D$, on the OAM states of the field scattered by a square array of $7 \times 7$ 
  spheres. The nearest neighbor distance is $870$~nm and the incident beam is a Gaussian
  ($\ell=0$) with beam waist $w_0=50$~$\mu$m at focus. (a) The azimuthal harmonics of the 
  distribution are defined as in Fig.~\ref{fig_Scat}. The changing distribution of 
  power ($P_{sca}$) amongst the OAM states ($l_z$) of the scattered field is shown in (b).
  (c) A cartoon schematic of the described setup.
  }
\end{figure}

In Fig.~\ref{fig:array_angle} we plot the OAM states of the field scattered by a 
square array of $7 \times 7$ identical spheres with a nearest neighbor distance 
of $870$~nm as the collection angle of the detector, $\theta_D$, is increased. 
The incident beam has a waist of $w_0=50$~$\mu$m at focus. 
\replaced[id=dm]{As the detection angle is increased, we observe four regimes
for the distribution of the scattered power amongst the OAM states. First, for 
$0<\theta_D< 10^\circ$, we observe the single OAM state $l_z=0$. Then, for 
$10<\theta_D< 30^\circ$ the OAM states}{
From left to right the detection angle is $5^\circ$, $20^\circ$ and $50^\circ$: 
with $5^\circ$ detection angle we observe a single OAM state, while the OAM 
states observed with $50^\circ$} have a periodicity $4$ originating from the 
$4$-fold symmetry of the structure. \added[id=dm]{For angles $30<\theta_D < 60^\circ$ some
of the scattered power goes into the OAM states corresponding to the first
harmonic of the structural symmetry, $l_z=\pm 8$. At these larger angles the
distribution of power amongst the OAM states qualitatively resembles the 
azimuthal moments of the particle distribution. Finally, for $\theta_D>60$ there
is energy distributed across all the even-numbered OAM states.}

\begin{figure}[ht!]
\includegraphics[clip=false]{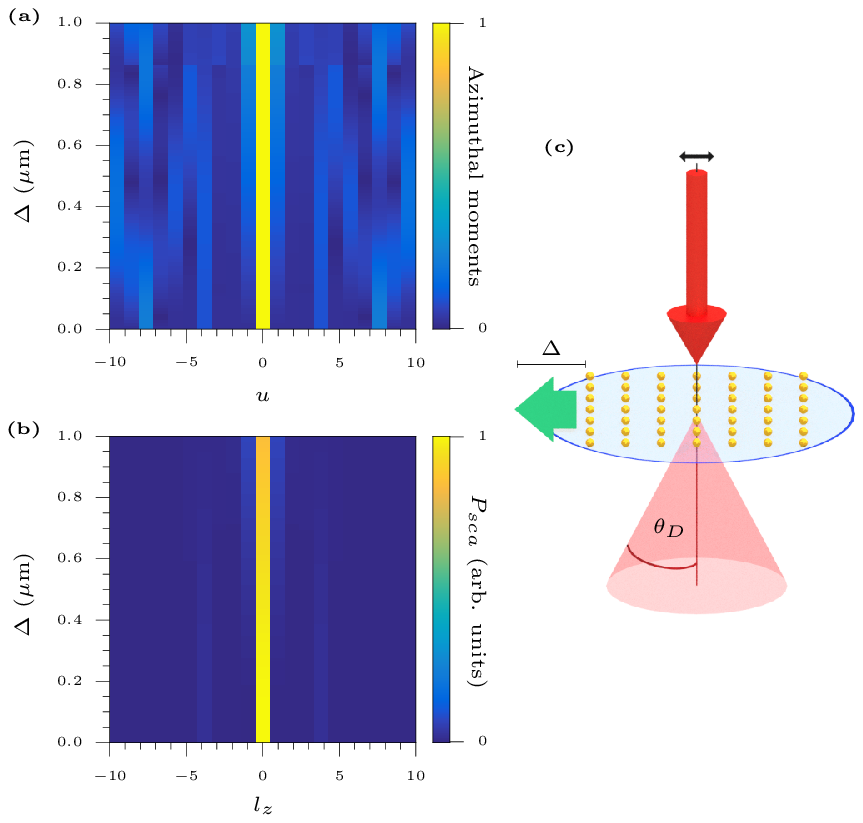}
  \caption{ \label{fig:array_axis} Effect of displacing the symmetry axis of a
  square array of particles away from the incident beam axis, by a distance $\Delta$,
  on the OAM states of the scattered field. The particle distribution and incident
  field are the same as in Fig.~\ref{fig:array_angle} and the detection angle
  $\theta_D=50^\circ$. (a) The azimuthal harmonics
  are defined with respect to the beam axis and vary with $\Delta$. (b) The array
  is displaced by up to $1\mu$m, and large relative variations in the power of the 
  OAM states with \replaced[id=dm]{$l_z=\pm 1$}{$l_z=\pm 1,\pm 2$} give a sensitivity to displacements of 
  $1\%$ of the beam waist. (c) A cartoon schematic of the described setup.
  }
\end{figure}

In Fig~\ref{fig:array_axis} we show the effect of displacing the \added[id=dm]{array 
of particles with respect to the} beam axis\deleted[id=dm]{ away from the 
symmetry axis}. We consider the same distribution as in Fig.~\ref{fig:array_angle} 
and a detection angle of $50^\circ$. 
\replaced[id=dm]{As the offset between the beam/symmetry axes increases, there
is a clearly visible effect on the OAM states of the scatted field, which now 
have strong peaks at $l_z=\pm 1$. This effect is also visible in the azimuthal
moments of the distribution which are defined with respect to the beam axis, 
and so also change as the particles are displaced.}
{An offset between the beam/symmetry axes of $500$~$\mu$m has a clearly visible 
effect on the OAM states of the scatted field, which now have strong peaks at 
$l_z=\pm2$ indicating an approximate $2$-fold symmetry. For an offset of 
$1$~$\mu$m the OAM states indicate an approximate $4$-fold symmetry.}
From the translation formulae, Eqs.~(\ref{eq:A3-1}-\ref{eq:transl}), we see that 
the spread of the scattered light power over $l_z$ increases with the 
displacement. It is therefore possible to determine the position of an axis of 
symmetry within the distribution of particles by finding the position, 
\replaced[id=dm]{with respect to}{of} the incident beam, that minimizes the 
spread of the scatted light power over $l_z$. 
Resolutions on the position of the symmetry axis of a few percent of the 
incident beam waist $w_0$ can be achieved with this method, depending on the 
ability to translate the position of the \added[id=dm]{particles, or } beam, 
and the precision in measuring the variation of the distribution of the 
scattered power over $l_z$.

\subsection{\label{sec:chiral} Chiral structures}
Finally, we investigate how \added[id=dm]{measuring} the OAM of the scattered 
fields\deleted[id=dm]{ therefore} provides a \replaced[id=dm]{method}{way} to
identify violations of symmetry \added[id=dm]{in distributions} under reflection 
with respect to \added[id=dm]{a plane parallel to the beam axis,} the $y=0$ 
plane.
\replaced[id=dm]{To do so, we study both two- and three-dimensional chiral distributions of 
particles~\footnote{Although no two-dimensional structure can be truly chiral,
they may still exhibit the properties associated with chiral 
structures~\cite{Barron04a}.}, specifically, particles arranged along a spiral and a 
helical path.}
{in planar chiral structures made by particles laying on a 
plane orthogonal to the propagation of the incident beam.}

We \added[id=dm]{start with the former, and} recall that, \added[id=dm]{for a 
distribution confined to the $xy$-plane,} a reflection with respect to the 
$y=0$ plane transforms a\deleted[id=dm]{ two-dimensional} right-handed structure 
into \replaced[id=dm]{its}{a} left-handed \replaced[id=dm]{counterpart}{structure}, 
and viceversa. 
\replaced[id=dm]{Such a}{The} reflection also transforms 
$m \rightarrow -m$, so the amplitude of the $u^\mathrm{th}$ harmonic of the 
distribution of one \added[id=dm]{type of} structure is the complex conjugate 
of the other.
\begin{figure}[ht!]
\includegraphics[clip=false]{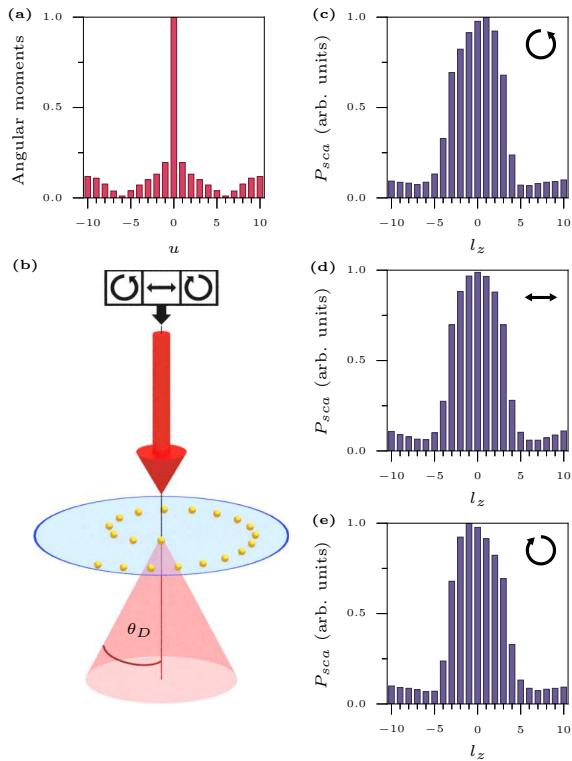}
  \caption{ \label{fig_spiral} Power of scattered light with defined OAM for a
  left-handed spiral consisting of $20$ nanospheres with a nearest neighbour 
  separation distance of $870$~nm. For all figures the incident beam is a 
  Gaussian, with beam waist $w_0=50$~$\mu$m, and the beam axis is perfectly 
  aligned with the symmetry axis of the distribution. (a) The azimuthal harmonics
  of the distribution are symmetric about $u=0$. The different OAM spectra of the scattered
  fields are shown for incident beams which are left circularly (c), linearly (d)
  and right circularly (e) polarized. (b) A cartoon schematic of the described setup.
  }
\end{figure}
Reflection symmetry breaking can be detected in the OAM states of the scattered 
light using incident beams with left and right circular polarization, as the 
power scattered into an OAM state depends on different azimuthal harmonics of 
the distributions.
This effect is most easily observed with wide detection angles. However, the 
detection does not distinguish between left-handed and right handed structures 
as the difference is only \added[id=dm]{encoded} in the phase of the azimuthal 
\replaced[id=dm]{components of the scattered light}{amplitudes}.
This is shown in Fig~\ref{fig_spiral}, where the chirality of a\deleted[id=dm]
{left-handed} spiral \added[id=dm]{distribution}, consisting of $20$ nanospheres, 
is evident from the difference in the OAM states of the scattered field for 
\added[id=dm]{incident} beams with \replaced[id=dm]{circular and linear}{left 
circular, linear and right circular} polarization.
With incident field polarization $\hat{e}_{\pm 1}$, the scattered light power 
has maxima for $l_z= \pm 1$ as these OAM values are originated by azimuthal 
harmonics of the multipole-multipole distributions with $u=-1,0,1$, which are 
the largest components of the distributions, as can be seen from the plot of 
the azimuthal moments.
\added[id=dm]{This method enables the identification of the presence of 
two-dimensional chiral structures but not their handedness, as this is encoded 
in the relative phases of the structures and this information is lost in the 
scattering process. Hence, equivalent results are obtained for both left and
right handed spirals.}

\begin{figure}[ht!]
\includegraphics[clip=false]{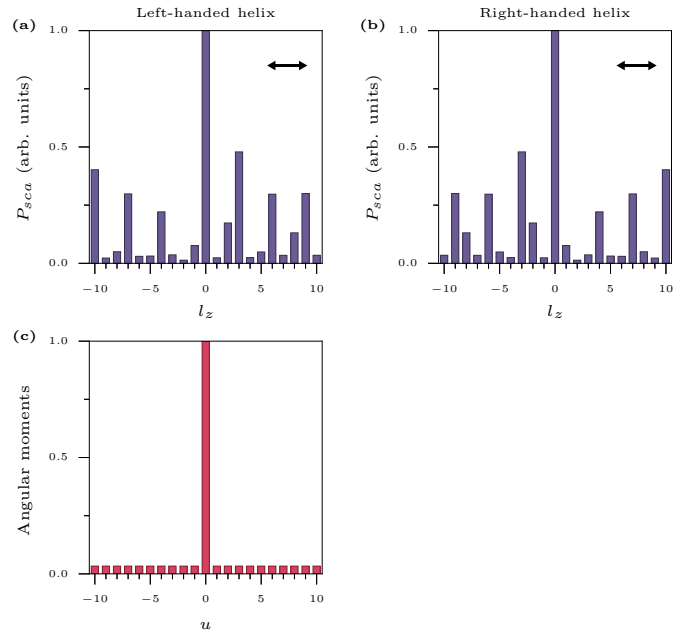}
  \caption{ \label{fig_helix} Power of scattered light with defined OAM for a 
  left-handed helix (a) and a right-handed helix (b) constructed
  by $30$ nanospheres with a pitch of $2.67$~$\mu$m and radius of 
  $10$~$\mu$m. For all figures the incident beam is a linearly polarized 
  Gaussian beam with beam waist $w_0=50$~$\mu$m. For helixes the 
  polarization of the incident beam is important and similar results can be 
  obtained with circular polarization (not shown). (c) The azimuthal moments
  of the mirror image structures are identical.
  }
\end{figure}

\replaced[id=dm]{In three dimensions, a}{Three dimensional} right-handed helix 
\replaced[id=dm]{is}{are instead} transformed into a left-handed helix by 
\replaced[id=dm]{an}{the} inversion \added[id=dm]{operation}. In this case, the 
phase difference between incident fields on different particles of the structure 
depends on the sign of $\ell$ and so we can distinguish \replaced[id=dm]{between}
{right-handed structures from left-handed} structures \added[id=dm]{with opposite 
handedness}. This is shown in Fig.~\ref{fig_helix} where, for a linearly polarized
incident Gaussian beam, the OAM \added[id=dm]{of the scattered field} clearly 
distinguishes between a left-handed and right-handed heli\replaced[id=dm]{cal structure}{x} 
\replaced[id=dm]{comprised of}{made by} $30$ nanospheres.
\replaced[id=dm]{Here}{In this case} the polarization of the incident beam is 
not \added[id=dm]{as} important and similar results are obtained with circular 
polarization (not shown). We note that the azimuthal moments, 
\replaced[id=dm]{and also}{as} the field scattered in the forward direction 
(not shown), do not capture the difference between a helix and a random structure
because the azimuthal moments are independent of \replaced[id=dm]{$z$}{the $z$ direction}, 
where the difference between a helix and random structure appears.

Also for these chiral structures a displacement \added[id=dm]{of the particles, with 
respect to the beam axis,} of $1\%$ of the beam waist is clearly detectable by 
observing variations in the power distribution of the scattered light over the 
OAM states (not shown).

\section{\label{sec:theory} Theory}
The total angular momentum of light  consists of a spin part and an orbital 
part that are distinct and physically meaningful but that are not themselves 
true angular momenta~\cite{barnett16}. Only the total angular momentum and its 
component along the $z$ axis are always well defined and conserved in 
light-matter interaction processes that are invariant under rotation with 
respect to a specific point and around $z$. This is the case only if the 
\added[id=dm]{position of the} scattering particle is centered on the axis of the incident beam. 
For both off-axis spheres and non-spherical particles the scattering process is 
not rotationally invariant and, therefore, the total angular momentum is not 
conserved. 

In the far field, it is straightforward to relate variations of the $z$ 
component of the total angular momentum to variations of the $z$ component of 
the OAM and the polarization, or spin angular momentum.
Therefore, the lack of rotational invariance provides the opportunity to 
understand properties of the particles' distributions by measuring, in the far 
field, variations of the OAM of the scattered field with respect to the OAM of 
the incident field. 
By expanding incident and scattered fields in terms of electric and magnetic 
multipoles~\cite{jackson99}, we derive a theory for quasi-monochromatic fields 
that allows us to determine the OAM of the scattered field in the far field 
zone as a function of the OAM of the incident beam and the electromagnetic 
properties, and spatial distributions, of particles. 
These are modeled by considering the spatial distributions of the 
multipole-multipole transition coefficients that at each point relate the 
amplitudes of the multipoles of the scattering fields to those of the incident 
fields.

We start with the Maxwell equations for electromagnetic waves with a harmonic 
time dependence, $\mathrm{e}^{-i\omega t}$:
\begin{eqnarray}
  \mathbf{E} = \frac{i}{k}\sqrt{\frac{\mu_r}{\varepsilon_r}}
    \left(\nabla\times\mathbf{H}\right); \;
  \mathbf{H} = -\frac{i}{k}\sqrt{\frac{\varepsilon_r}{\mu_r}}
    \left(\nabla\times\mathbf{E}\right)
\end{eqnarray}
where we work in SI units and have rescaled the electric and magnetic fields,
$\mathbf{E} = \sqrt{\varepsilon_0} \, \mathbf{E}^{\prime}$ and
$\mathbf{H} = \sqrt{\mu_0} \, \mathbf{H}^{\prime}$, 
where $\varepsilon_0, \mu_0$ are the dielectric permitivitty and magnetic 
permeability of vacuum, respectively, with $k$ the wavenumber in the medium.
We also adopt a compact notation with six-dimensional vectors for the 
electromagnetic fields, 
$\mathcal{F}(\mathbf{r})=[\mathbf{E}(\mathbf{r})^\tr,\mathbf{H}(\mathbf{r})^\tr]^\tr$, 
where the first three components are electric, the last three magnetic and $\tr$ 
stands for the transpose~\cite{papoff11a,mcarthur17b,mcarthur17c}.


We start by considering an incident plane wave propagating in the direction 
specified by the wave vector $\mathbf{k}_0$:
\begin{equation}
  \mathcal{F}^0 (\mathbf{r}) = \big[{\mathbf{E}^0}^\tr, {\mathbf{H}^0}^\tr \big]^\tr 
      \exp{(i \mathbf{k}_0 \cdot \mathbf{r})}
\end{equation}
where we define the \added[id=dm]{Cartesian} unit vector $\hat{\mathbf{r}}=
(\sin{\theta}\cos{\varphi},\sin{\theta}\sin{\varphi},\cos{\theta})$ and the 
wave vector $\mathbf{k}_0= \left| \mathbf{k}_0 \right| 
(\sin{\theta_0}\cos{\varphi_0},\sin{\theta_0}\sin{\varphi_0},\cos{\theta_0})$ 
in terms of angular spherical coordinates, and $\mathbf{E}^0, \mathbf{H}^0$ are 
the vector amplitudes of the electric and magnetic fields.

Expanding the exponential term $\exp{(i \mathbf{k}_0 \cdot \mathbf{r})}$
using the vector spherical harmonic basis~\cite{jackson99}, adopting Einstein's 
convention of summing over repeated indices, the plane wave above can be 
written as:
\begin{equation}
  \mathcal{F}^0 (\mathbf{r}) = 
    \Big[a^H_{j^{\prime}m^{\prime}} \mathcal{J}^H_{j^{\prime}m^{\prime}}
    (\mathbf{r}) + a^E_{j^{\prime}m^{\prime}} 
    \mathcal{J}^E_{j^{\prime}m^{\prime}}(\mathbf{r})\Big],
\end{equation}
where $j^{\prime}$ is the index of the \emph{total} angular momentum, 
$m^{\prime}$ is the index of the total angular momentum projected along $z$, 
and $a^H_{j^{\prime}m^{\prime}}$ and $a^E_{j^{\prime}m^{\prime}}$ are beam 
expansion coefficients, given by 
\begin{subequations} 
\begin{eqnarray}
  a^H_{j^{\prime}m^{\prime}} &=& 4\pi i^{j^{\prime}} 
    \mathbf{m}^*_{j^{\prime}m^{\prime}}(\theta_0, \varphi_0) \cdot 
    \mathbf{E}^0, \label{eq:pw_bec_h} \\
  a^E_{j^{\prime}m^{\prime}} &=& \sqrt{\frac{\mu_r}{\varepsilon_r}}4\pi i^{j^{\prime}+1} 
    \mathbf{m}^*_{j^{\prime}m^{\prime}}(\theta_0, \varphi_0) \cdot 
    \mathbf{H}^0. \label{eq:pw_bec_e}
\end{eqnarray}
\end{subequations}
The vector spherical harmonic $\mathbf{m}_{j^{\prime}m^{\prime}}(\theta_0,
\varphi_0)$ is defined as in Appendix~\ref{app:A1}.
$\mathcal{J}^E_{j^{\prime}m^{\prime}}(\mathbf{r})$ and 
$\mathcal{J}^H_{j^{\prime}m^{\prime}}(\mathbf{r})$ are regular electric and 
magnetic spherical multipoles -- i.e. vector spherical standing waves --
centered in $\mathbf{r}=0$, and given in Eqs.~(\ref{eq:sph_multi_H}-\ref{eq:sph_multi_E})
The incident field is the field that would exist without a scatterer and 
therefore includes both incoming and outgoing parts. As it should be finite 
everywhere $\mathcal{J}^{E, H}_{j^{\prime}m^{\prime}}$ are written in terms of 
regular Bessel functions.

Note that the electric and magnetic multipoles are exact solutions of Maxwell's 
equations and it is straightforward to verify that they are eigenfunctions of 
the operators of the total angular momentum, $\hat{J}^2$ and $\hat{J}_z$: 
$\hat{J}^2 \mathcal{J}^{E, H}_{j^{\prime}m^{\prime}}=j^{\prime}
(j^{\prime}+1)\mathcal{J}^{E, H}_{j^{\prime}m^{\prime}}$ and 
$\hat{J}_z \mathcal{J}^{E, H}_{j^{\prime}m^{\prime}}=m^{\prime} 
\mathcal{J}^{E, H}_{j^{\prime}m^{\prime}}$.
%

We first consider the field scattered by a non-spherical particle with center 
at $\mathbf{r}=0$. Assuming the scattering problem is linear~\cite{Kiselev14},
we can relate the beam expansion coefficients for the scattered waves to those 
of the incident waves via
\begin{subequations} 
\begin{eqnarray}
  a^{H, s}_{j m} &=& a^E_{j^{\prime}m^{\prime}} T^{EH}_{j^{\prime}m^{\prime} j m} 
    + a^H_{j^{\prime}m^{\prime}} T^{HH}_{j^{\prime}m^{\prime} j m} \\
  a^{E, s}_{j m} &=& a^E_{j^{\prime}m^{\prime}} T^{EE}_{j^{\prime}m^{\prime} j m}
    + a^H_{j^{\prime}m^{\prime}} T^{HE}_{j^{\prime}m^{\prime} j m}, 
\end{eqnarray}
\end{subequations}
where $T^{EH},T^{HH},T^{EE},T^{HE}$ are partitions of the $T-matrix$ which 
account for the dependence of the multipoles in the expansion of the 
scattered field on the multipoles in the expansion of the incident 
field~\cite{mishchenko2000,mishchenko04a}.
In general the scattering process mixes angular momenta~\cite{mishchenko96a}; 
when this happens, the sum over $j^{\prime}m^{\prime}$ involves terms different 
from $jm$.
The first superscript \added[id=dm]{index of the matrix partitions} indicates 
the type of multipole in the incident field expansion that produces the type of 
multipole response in the scattered field expansion indicated by the second 
superscript\added[id=dm]{; electric ($E$) or magnetic ($H$)}.
These partitions depend only on the wavelength, the size and shape of the 
particle and the permeability and permitivitty of the internal and external 
media~\cite{waterman71a,doicu06a,waterman07a}. 

Thus we can write the scattered field as
\begin{equation}
  \mathcal{F}^{s}(\mathbf{r}) = \left[ a^{H, s}_{j m}
    \mathcal{S}^H_{jm}(\mathbf{r}) +  a^{E, s}_{j m}
    \mathcal{S}^E_{jm}(\mathbf{r})\right]
\label{eq:sscat-nsp}
\end{equation}
where we expand the scattered field in terms of radiating electric and magnetic 
multipoles, $\mathcal{S}^E_{jm}$ and $\mathcal{S}^H_{jm}$. These have the same 
form as (\ref{eq:sph_multi_H}-\ref{eq:sph_multi_E}) but with the Bessel 
functions replaced by spherical Hankel functions that are singular at the 
origin, and are also eigenfunctions of $\hat{J}^2$ and $\hat{J}_z$, as above. 
Once the $T$-matrix is \replaced[id=dm]{constructed}{calculated}, scattering 
from non-spherical particles can be 
calculated efficiently without having to perform surface or volume integrals.

%
We note that in Eq.~(\ref{eq:sscat-nsp}) it is essential that the beam 
expansion coefficients are calculated with respect to the center of the 
scattering particle. If we now allow the particle to be centered at some 
arbitrary position $\mathbf{r}_t$ we can write the scattered field as
\begin{equation}
  \mathcal{F}^{s}_t(\mathbf{r}) = \left[a^{H, s}_{j m}
    \mathcal{S}^H_{jm}(\mathbf{r} - \mathbf{r}_t) +  a^{E, s}_{j m}
    \mathcal{S}^E_{jm}(\mathbf{r} - \mathbf{r}_t)\right], \label{eq:sscat-nsp2}
\end{equation}
where $j, m$ are associated with the angular momenta in the center of the 
particle $\mathbf{r}_t$. This means that 
$\hat{J}^2_t {S}^E_{jm}(\mathbf{r} - \mathbf{r}_t) = 
j(j+1) {S}^E_{jm}(\mathbf{r} - \mathbf{r}_t)$ and 
$\hat{J}_{tz}{S}^E_{jm}(\mathbf{r} - \mathbf{r}_t) = 
m {S}^E_{jm}(\mathbf{r} - \mathbf{r}_t)$, where $\hat{J}_t$
is the total angular momentum  operator with respect to $\mathbf{r}_t$ and
$\hat{J}_{tz}$ its component along $z$. However, 
$\hat{J}^2 {S}^E_{jm}(\mathbf{r} - \mathbf{r}_t) \ne 
j(j+1) {S}^E_{jm}(\mathbf{r} - \mathbf{r}_t)$ and 
$\hat{J}_z{S}^E_{jm}(\mathbf{r} - \mathbf{r}_t) \ne 
m {S}^E_{jm}(\mathbf{r} - \mathbf{r}_t)$.

For a dilute solution of particles, with inter-particle distances such that 
multiple scattering can be neglected~\cite{berg08a}, the total scattered field 
can then be found by summing the individual fields due to each particle, i.e. 
by summing the fields given by Eq.~(\ref{eq:sscat-nsp2}) with beam expansion 
coefficients calculated with respect to the center of each 
particle~\cite{gutierrez-cuevas18a},
\begin{eqnarray}
  \mathcal{F}^{s}(\mathbf{r}) = \sum_t \mathcal{F}^{s}_t(\mathbf{r}) &=& 
    \Big[ a^{H, s}_{j m} (\mathbf{r}_t) 
    \mathcal{S}^H_{jm}(\mathbf{r} - \mathbf{r}_t)\nonumber \\
    &&+ a^{E, s}_{j m} (\mathbf{r}_t)
    \mathcal{S}^E_{jm}(\mathbf{r} - \mathbf{r}_t)\Big], \label{eq:sum-scat-nsp2}
\end{eqnarray}
where $a^{H, s}_{j m}(\mathbf{r}_t)$ and $a^{E, s}_{j m}(\mathbf{r}_t)$ are the 
coefficients for the particle at $\mathbf{r}_t$.

%
When the incident beam consists of an arbitrary superposition of plane waves, 
at the position of each particle we write it as:
\begin{eqnarray}
  \mathbf{E}(\mathbf{r}_t) = \int \tilde{\mathbf{E}}(\theta_k, \varphi_k)
    \exp{\left[i \mathbf{k}(\theta_k, \varphi_k) \cdot \mathbf{r}_t \right]}\,
    d \theta_k d \varphi_k,\\
  \mathbf{H}(\mathbf{r}_t) = \int \tilde{\mathbf{H}}(\theta_k, \varphi_k)
    \exp{\left[i \mathbf{k}(\theta_k, \varphi_k) \cdot \mathbf{r}_t \right]}\,
    d \theta_k d \varphi_k.
\end{eqnarray}
The corresponding beam expansion coefficients are scalar functions of the 
positions of the particles' centers
\begin{subequations} 
\begin{eqnarray}
  A^H_{j^{\prime}m^{\prime}}(\mathbf{r}_t) &=& 4\pi i^{j^{\prime}}  \int 
    {\bf m}^*_{j^{\prime}m^{\prime}}(\theta_k, \varphi_k) \cdot 
    \tilde{ {\bf E}}(\theta_k, \varphi_k) \nonumber \\
   && \times \exp{\left[i{\bf k}(\theta_k,\varphi_k)\cdot\mathbf{r}_t \right]}\,
      d \theta_k d \varphi_k, \label{eqn:genAH}\\
  A^E_{j^{\prime}m^{\prime}}(\mathbf{r}_t) &=& \sqrt{\frac{\mu_r}{\varepsilon_r}}
    4\pi i^{j^{\prime}+1} \int{\bf m}^*_{j^{\prime}m^{\prime}}(\theta_k,\varphi_k)
    \cdot \tilde{ {\bf H}}(\theta_k, \varphi_k) \nonumber \\
   && \times \exp{\left[i {\bf k}(\theta_k, \varphi_k) \cdot \mathbf{r}_t \right] }\,
    d \theta_k d \varphi_k. \label{eqn:genAE}
\end{eqnarray}
\end{subequations}
\deleted[id=dm]{where ${\bf m}^*_{j^{\prime}m^{\prime}}(\theta_k, \varphi_k)$ 
is the vector spherical harmonic defined in Eq.~(\ref{eq:sph_m}).}

The field scattered by the distribution of particles is then
\begin{eqnarray}
  \mathcal{F}^{s}(\mathbf{r}) = \sum_t  \mathcal{F}^{s}_t(\mathbf{r}) &=& 
    \Big[ A^{H, s}_{j m} (\mathbf{r}_t) \mathcal{S}^H_{jm}(\mathbf{r} - \mathbf{r}_t) \nonumber \\
    &&+ A^{E, s}_{j m} (\mathbf{r}_t) \mathcal{S}^E_{jm}(\mathbf{r} - \mathbf{r}_t)\Big] .
\label{eq:sscat-nsp2b}
\end{eqnarray}
where the beam expansion coefficients of the scattered waves due to a particle 
centered on 
$\mathbf{r}_t$, $A^{H, s}_{j m} (\mathbf{r}_t), A^{E, s}_{j m} (\mathbf{r}_t)$, 
are related to the beam expansion coefficients of the incident waves via the 
$T$-matrix as before
\begin{subequations} 
\begin{eqnarray}
  \hspace*{-2em} A^{H, s}_{j m} (\mathbf{r}_t) &=& A^E_{j^{\prime}m^{\prime}}(\mathbf{r}_t) 
    T^{EH,t}_{j^{\prime}m^{\prime} j m}+ A^H_{j^{\prime}m^{\prime}}(\mathbf{r}_t)
    T^{HH,t}_{j^{\prime}m^{\prime} j m}  \\
  \hspace*{-2em} A^{E, s}_{j m} (\mathbf{r}_t) &=& A^E_{j^{\prime}m^{\prime}}(\mathbf{r}_t)
    T^{EE,t}_{j^{\prime}m^{\prime} j m} +A^H_{j^{\prime}m^{\prime}}(\mathbf{r}_t) 
    T^{HE,t}_{j^{\prime}m^{\prime} j m}
\end{eqnarray}
\end{subequations}
where the indices $j^{\prime}, m^{\prime}, j, m$ are associated with the 
angular momenta in the center of the particle.

%
In order to evaluate the angular momentum with respect to the reference frame 
of the beam, centered in $\mathbf{r}=\mathbf{0}$, we express 
Eq.~(\ref{eq:sscat-nsp2b}) in terms of products of functions that depend only 
on $\mathbf{r}$ or on $\mathbf{r}_t$. This is done by applying the asymptotic 
form of the translation formulae and the Jacobi-Anger identity, which are valid
in the far field region, see Eq.~(\ref{eq:transl}) in Appendix \ref{app:A3}. 
In the far field region, the field scattered by all the particles is
%
\begin{widetext}
\begin{equation}
  \begin{array}{lcl}
    \mathcal{F}^{s\infty}(\theta,\varphi) & = & \exp{(-ik \hat{\mathbf{r}}
    \cdot \mathbf{r}_t)} \left[ A^{H, t}_{j m} (\mathbf{r}_t)
    \mathcal{S}^{H\infty}_{jm}(\theta,\varphi)+ A^{E, t}_{j m} (\mathbf{r}_t) 
    \mathcal{S}^{E\infty}_{jm}(\theta,\varphi)\right] \\[2ex]
    &=& \exp{(-ik |\mathbf{r}_t| \cos{ \theta} \cos{\theta_t})}
    \displaystyle{\sum_{n=-\infty}^{\infty}} (-i)^n \exp{(-i n\varphi_t)} 
    J_n(k |\mathbf{r}_t| \sin{ \theta} \sin{\theta_t}) \\[2ex]
    && \times  \left[ A^{H, t}_{j m} (\mathbf{r}_t)
    \mathcal{S}^{H\infty}_{jm}(\theta,\varphi) + A^{E, t}_{j m} (\mathbf{r}_t)
    \mathcal{S}^{E\infty}_{jm}(\theta,\varphi)\right] \exp{(i n\varphi)}.
  \end{array} \label{eq:ff-sscat-jsphere-nsp}
\end{equation}
\end{widetext}
The functions $\mathcal{S}_{jm}^{E\infty}, \mathcal{S}_{jm}^{H\infty}$ describe 
the scattering waves in the asymptotic limit as $|\mathbf{r}|\rightarrow\infty$,
see Eqs.~(\ref{eq:svwf-ff-E},\ref{eq:svwf-ff-H}) in Appendix \ref{app:A3}, and 
$J_n(\cdot)$ is a Bessel function of the first kind of order $n$.
The indices $j,m$ are now referring to the center of the incident beam, while 
the index $n$ gives an extra contribution to $\hat{J}_z$ arising from the fact 
that fields scattered by particles displaced from the $z$-axis have an 
additional azimuthal dependence. This result comes from the translation 
formulae: note that 
$\hat{J}_z \mathcal{S}_{jm}^{E\infty} \exp{(i n\varphi) = (m+n) 
\mathcal{S}_{jm}^{E\infty} \exp{(i n\varphi)}}$.

%
In order to understand how the scattered field is affected by both the incident 
field and the distribution of particles, and hence make analytical predictions, 
we introduce the density of multipole-multipole transitions. 
This density is made up by a sum of delta functions that depend on the 
continuous variable $\mathbf{r}^{\prime}$ and on the position of the particles' 
centers, $\mathbf{r}_t$; the volume integral of this density over 
$\mathbf{r}^{\prime}$ gives the distribution of multipoles centered in 
$\mathbf{r}_t$  associated with the corresponding distribution of particles.
The density is
%
%
\begin{eqnarray}
  D^{AB}_{j^{\prime}m^{\prime} j m}(\mathbf{r}^{\prime}) &=& 
    {\rho^{\prime}}^{-1} \delta(\rho^{\prime}-\rho_t) \,
    \delta(\varphi^{\prime}-\varphi_t) \, \delta(z^{\prime}-z_t) \, \nonumber \\
    && \times T^{AB,t}_{j^{\prime}m^{\prime} j m} \\[2ex]
  &=& D^{AB}_{j^{\prime}m^{\prime} j m u}(\rho^{\prime},z^{\prime}) \exp{(iu \varphi^{\prime})},     
\end{eqnarray}
where $z_t=|\mathbf{r}_t| \cos{\theta_t}$, and 
$\rho_t=|\mathbf{r}_t| \sin{\theta_t}$, $\delta(\rho^{\prime}-\rho_t)$, 
$\delta(z^{\prime}-z_t)$ and 
$\delta(\varphi^{\prime}-\varphi_t)=(2\pi)^{-1} \exp{(-iu\varphi_t)} 
\exp{(iu\varphi^{\prime})}$
(with sum over the repeated index $u \in \mathbb{Z}$) are Dirac delta functions, 
and the multi-index superscripts $A = E, H$ and $B = E, H$.
The $u^\mathrm{th}$ azimuthal Fourier component of 
$D^{AB}_{j^{\prime}m^{\prime} j m}(\mathbf{r}^{\prime})$ is
\begin{eqnarray}
  D^{AB}_{j^{\prime}m^{\prime} j m u}(\rho^{\prime},z^{\prime}) &=& (2\pi\rho^{\prime})^{-1} 
    \delta(\rho^{\prime}-\rho_t)\delta(z^{\prime}-z_t) \nonumber \\
    && \times T^{AB,t}_{j^{\prime}m^{\prime} j m} \exp{(-iu\varphi_t)}.
\end{eqnarray}
For distributions of identical particles, the volume integral of 
$D^{AB}_{j^{\prime}m^{\prime} j m}(\mathbf{r}^{\prime})$ over the space 
occupied by the distribution is proportional to the $u^\mathrm{th}$ azimuthal 
moment of the distribution.

These densities are characteristic of the whole set of particles and allow us 
to replace sums over the number of particles with integrals over the volume 
occupied by the particles in the calculation of the scattered field. Note that  
the smallest scales $\lambda$ and $R_t$ affect only the terms $T^{AB,t}$, 
while the remaining factors in the densities take into account spatial 
variations at scales ranging from the shorter inter-particle distance to the 
macroscopic scales $\mathcal{L}_\perp$ and $\mathcal{L}_\parallel$.

Using these distributions and considering the beam coefficients' dependence 
upon the azimuthal angle $\varphi^{\prime}$, 
$A^{E, H}_{j^{\prime}m^{\prime}}(\mathbf{r}^{\prime}) = 
A^{E/H}_{j^{\prime}m^{\prime} q}(\rho^{\prime},z^{\prime}) 
\exp{(i q \varphi^{\prime})}$, 
the spatial integration gives the selection rule $n=u+q$ that reveals the 
connection between the scattered field and the spatial properties of the 
distributions and incident field. Note that for an arbitrary incident field, 
the relation between the optical angular momentum of the incident beam, $\ell$, 
and $q$ can be calculated using translation-addition formulae. 
As the scattered field becomes
%
\begin{eqnarray}
  \mathcal{F}^{s\infty}(\theta, \varphi) &=& \Big[\left(\Pi_{jmqu}^{EH}(\theta)
    + \Pi_{jmqu}^{HH}(\theta)\right)\mathcal{S}_{jm}^{H \infty}(\theta,\varphi)
    \nonumber \\
  &&+ \left(\Pi_{jmqu}^{EE}(\theta) + \Pi_{jmqu}^{HE}(\theta)\right)
     \mathcal{S}_{jm}^{E \infty}(\theta, \varphi)\Big] \nonumber \\
  && \times \exp{[i (u+q)\varphi]},
 \label{eq:ff-sscat-density2-nsp}
\end{eqnarray}
where we define the complex effective scattering amplitudes
%
\begin{widetext}
\begin{eqnarray}
   \Pi_{jmqu}^{AB}(\theta) &=& 
      (-i)^{u+q}  \int {A^{A}}_{j^{\prime}m^{\prime} q}(\rho^{\prime},z^{\prime})
      D^{AB}_{j^{\prime}m^{\prime} j m u}(\rho^{\prime},z^{\prime})\exp{(-ik z^{\prime} \cos{\theta})} 
      J_{u+q}( k \rho^{\prime} \sin{\theta}) dV  \nonumber \\
   &=&(-i)^{u+q}  {A^{A}}_{j^{\prime}m^{\prime} q}(\rho_t, z_t) T^{AB,t}_{j^{\prime}m^{\prime} j m} \exp{(-i u \varphi_t)}
      \exp{(-ik z_t \cos{\theta} )} J_{u+q}( k \rho_t \sin{\theta}).
\end{eqnarray}
\end{widetext}
Eq.~(\ref{eq:ff-sscat-density2-nsp}) relates the total angular momenta of the 
scattered light, with respect to the beam center, and its component along the 
beam's axis to the macroscopic properties of the distribution of particles 
and the incident beam. 
This can be seen in terms of the operator $\hat{J}_z$ from the identity 
$\hat{J}_z \mathcal{S}_{jm}^{E\infty} \exp{[i (u+q)\varphi]} = 
(m+u+q) \mathcal{S}_{jm}^{E\infty} \exp{[i (u+q)\varphi]}$. 
This equation is general and can be applied to \emph{any} set of dilute 
(non-interacting) particles and \emph{any} type of incident field.
Furthermore, it shows that the far field scattering has the same form as the 
single particle scattering, Eq.~(\ref{eq:sscat-nsp}), but with amplitude terms 
dependent on the angle $\theta$ that are moments of the distributions of the 
multipole-multipole transitions. These terms also contain an extra contribution 
to $\hat{L}_z$ -- i.e. the dependence on $\varphi$ -- that depends on the 
spatial distribution of the multipole-multipole transitions and the beam 
expansion coefficients.

From Eq.~(\ref{eq:ff-sscat-density2-nsp}) we can derive the following 
properties that apply to distributions of particles with any shape and for any 
incident field, as long as multiple scattering can be neglected:
\begin{itemize}
\item
{\bf For each spatial harmonic $\bm{u}$ of the multipole-multipole distributions and
harmonic $\bm{q}$ of the beam expansion coefficient, the  multipolar waves 
$\bm{\mathcal{S}_{jm}^{H \infty}}$ and $\bm{\mathcal{S}_{jm}^{E \infty}}$, scattered by 
every particle, add coherently to form a scattered wave with $\bm{j_z = u+q+m}$.}
\item
{\bf For a scattered wave with $\bm{j_z = u+q+m}$, the component of orbital angular 
momentum along the beam axis, $\bm{l_z}$, takes three values: $\bm{l_z= j_z-s_z}$, with 
$\bm{s_z=0}$ for polarization $\mathbf{\hat{z}}$, and $\bm{s_z = \pm 1}$ for polarization 
$\mathbf{\hat{e}}_{\mp}$.}
\item
{\bf For $\bm{\theta=0,\pi}$, i.e. in the forward and backward directions, only the 
terms with $\bm{u=-q}$ and either $\mathbf{\hat{e}}_+$ and $\bm{m= 1}$, or 
$\mathbf{\hat{e}}_-$ and $\bm{m=-1}$, do not vanish, as can be seen by the 
dependence on the angular variables of $\bm{\Pi_{jmqu}^{AB}}$, 
$\bm{\mathcal{S}_{jm}^{E \infty}}$ and $\bm{\mathcal{S}_{jm}^{H \infty}}$. For these 
terms $\bm{l_z=0}$, however the two conditions above are more restrictive than $\bm{l_z=0}$.}
\end{itemize}
These properties are most useful when the $q$-index of the incident field has a 
single value: this is the case for the Laguerre-Gaussian paraxial fields in the 
local plane wave approximation, discussed in the following section, where 
$q=\ell$.

\section{\label{sec:PWA} Local plane wave approximation}
It is interesting to consider the case of spheres because the theory becomes 
fully analytical and a very large number of experiments are performed with 
spheres. As a consequence of the spherical symmetry, both $\hat{J}$ and 
$\hat{J}_z$ are conserved and, therefore, $T^{EH}=T^{HE}=0$, $T^{EE},T^{HH}$ 
are diagonal in $j,j^{\prime}$ and $m,m^{\prime}$ and their elements are the 
Mie coefficients, that are independent from $\hat{J}_z$, which can be 
calculated analytically~\cite{Mie1908}.

If the incident field is now an arbitrary Laguerre-Gaussian paraxial beam 
propagating along the $z$ axis, the rescaled fields are, in the units used in 
this paper,
%
\begin{eqnarray}
  \mathbf{E}^0(\mathbf{r}) &=& \Big[\mathbf{\hat{e}}^0 u_{\ell p}(\mathbf{r}) 
    + \mathbf{\hat{z}} \frac{i}{k}\p_x u_{\ell p}(\mathbf{r})\Big] \exp{(ikz)}, \\
  \mathbf{H}^0(\mathbf{r}) &=& \sqrt{\epsilon_r/\mu_r} \Big[\left(\mathbf{\hat{z}}
    \times \mathbf{\hat{e}}^0\right) u_{\ell p}(\mathbf{r}) \nonumber \\
   && + \mathbf{\hat{z}} \frac{i}{k}\p_y u_{\ell p}(\mathbf{r})\Big]\exp{(ikz)} ,
\end{eqnarray}
where $\mathbf{\hat{e}}^0$ is a polarization vector which satisfies
$\mathbf{\hat{z}}\cdot \mathbf{\hat{e}}^0=0$ and 
$u_{\ell p}(\mathbf{r})=LG_{\ell p}(\rho, z) \exp{(i\ell \varphi)}$ is a 
Laguerre-Gaussian amplitude distribution~\cite{siegman86a}, with radial index 
$p$ and azimuthal index $\ell$. Note that we have included a factor 
$i\omega\sqrt{\epsilon_0}$ in $LG_{\ell p}$ with respect to typical expressions 
from literature~\cite{allen99,haus1984}, where $\omega$ is the angular 
frequency of the field. 
Assuming that the largest dimension of the particles, $R_t \ll w_0, z_R$, with 
$w_0$ the beam waist of the Laguerre-Gaussian beam and, $z_R$ the corresponding 
Rayleigh parameter $z_R=k w_0^2/2$, then each sphere sees three waves 
propagating along $z$: one transverse and two (spurious) longitudinal, which do
not produce scattering as shown in Appendix \ref{app:A2}.
For this reason, only corrections to the paraxial fields of order $k^{-2}$ 
would affect the scattered fields; for the cases considered here, corrections 
of this order can be safely neglected.

From the plane wave expansion coefficients, Eqs.~(\ref{eq:pw_bec_h}-
\ref{eq:pw_bec_e}) in Appendix \ref{app:A2}, we find that the scattered field 
of a distribution of spherical particles in the far field limit is
%
\begin{eqnarray}
  \mathcal{F}^{s\infty}(\theta,\varphi) &=& \Big[\Pi_{j,-1,\ell u}^{HH}(\theta) 
      \mathcal{S}_{j,-1}^{H \infty}(\theta, \varphi) \nonumber \\
    && + \Pi_{j,1,\ell u}^{HH}(\theta) \mathcal{S}_{j,1}^{H \infty}(\theta, \varphi) \nonumber \\
    && + \Pi_{j,-1,\ell u}^{EE}(\theta) \mathcal{S}_{j,-1}^{E \infty}(\theta, \varphi) \nonumber \\
    && + \Pi_{j,1,\ell u}^{EE}(\theta) \mathcal{S}_{j,1}^{E \infty}(\theta, \varphi)\Big] \nonumber \\
    && \times \exp{[i (u+\ell )\varphi]},  \label{eq:ff-sscat-densityLG}
\end{eqnarray}
with complex scattering amplitudes
%
\begin{subequations} 
\begin{eqnarray}
  \Pi_{j,\pm1,\ell u}^{EE}(\theta) &=& i^{j+1-u-\ell }  \sqrt{2 \pi(2j+1)}
    \mathbf{\hat{e}}_{\mp} \cdot (\mathbf{\hat{z}} \times \mathbf{\hat{e}}^0) \label{eq:Pi_sphr_E}\\
    && \times \dsp\int \rho^{\prime} d\rho^{\prime} dz^{\prime} LG_{\ell p}(\rho^{\prime},z^{\prime})
    D^{EE}_{j,\pm 1,u}(\rho^{\prime},z^{\prime}) \nonumber \\
    && \times \exp{[ik z^{\prime} (1-\cos{\theta})]} J_{u+\ell }( k \rho^{\prime} \sin{\theta}),\nonumber\\
  \Pi_{j,\pm1,\ell u}^{HH}(\theta) &=& i^{j-u-\ell }  \sqrt{2 \pi(2j+1)}
    \mathbf{\hat{e}}_{\mp} \cdot \mathbf{\hat{e}}^0 \label{eq:Pi_sphr_H} \\
    && \times  \dsp\int \rho^{\prime} d\rho^{\prime} dz^{\prime} LG_{\ell p}(\rho^{\prime},z^{\prime})
    D^{HH}_{j,\pm 1,u}(\rho^{\prime},z^{\prime}) \nonumber \\
    && \times \exp{[ik z^{\prime} (1-\cos{\theta})]} J_{u+\ell }( k \rho^{\prime} \sin{\theta}).\nonumber
\end{eqnarray}
\end{subequations}
In Eqs.~(\ref{eq:Pi_sphr_E}-\ref{eq:Pi_sphr_H}) we have dropped for $D^{AA}$ 
the indexes corresponding to the incident fields because they are the same as 
those for the scattered field for the properties of the Mie coefficients 
mentioned before.

Evaluating Eqs.~(\ref{eq:Pi_sphr_E}-\ref{eq:Pi_sphr_H}) at $\theta=0$, the only 
terms that contribute to the scattered field can be written as
%
\begin{subequations} 
\begin{eqnarray}
  \Pi_{j,\pm1,\ell ,-\ell }^{EE}(0) &=& i^{j+2-u-\ell }  \sqrt{2\pi(2j+1)} \nonumber \\
    && \times \mathbf{\hat{e}}_{\mp} \cdot (\mathbf{\hat{z}}
    \times \mathbf{\hat{e}}^0) 
    (LG_{\ell p}^* ,  D^{EE}_{j,\pm1,-\ell }) \label{eq:pi_ee} \\
  \Pi_{j,\pm1,\ell ,-\ell }^{HH}(0) &=& i^{j+1-u-\ell } \sqrt{2 \pi(2j+1)} \nonumber \\ 
    && \times \mathbf{\hat{e}}_{\mp} \cdot \mathbf{\hat{e}}^0 
    (LG_{\ell p}^* , D^{HH}_{j,\pm1,-\ell }), \label{eq:pi_hh}
\end{eqnarray}
\end{subequations}
where we have introduced the notation 
$(f , g) = \int f^*(\rho,z) g(\rho,z) \rho d\rho dz $ to define an overlap 
integral over the cylindrical distribution volume, projected along the 
azimuthal angle.
An important feature of Eqs.~(\ref{eq:pi_ee}-\ref{eq:pi_hh}) is that the radial 
and axial coordinates of the particles, $\rho_t$ and $z_t$, affect the 
scattering amplitudes only through the slowly varying amplitude $LG_{\ell p}$ 
of the Laguerre-Gauss beam. As this amplitude does not vary significantly for 
$z>z_R$, most of the particles scatter in phase in the forward direction.
For $\theta=\pi$, i.e. the backward scattered field, Eqs.~(\ref{eq:pi_ee}-
\ref{eq:pi_hh}) are modified by adding a factor of $\exp{(i2 k z^{\prime})}$ 
inside the integrals.
When the longitudinal dimension of the distribution is significantly larger 
than the wavelength, $\mathcal{L}_\parallel\gg \lambda$, the stationary phase 
approximation shows that the dominant term in 
$\Pi_{j,\pm,\ell ,-\ell }^{EE/HH}(0)$ is the coefficient of the Fourier 
component of $D^{EE/HH}$ that varies along $z$ as $\exp{(-i2 k z^{\prime})}$.
In practice, this means that in the backward direction the scattering intensity 
is dominated by distributions periodic along $z$, of spatial period $\lambda/2$, 
or by particles with the same $z$ coordinate, whose scattered fields add in 
phase.

\section{\label{sec:conc} Conclusions}
In this paper we have presented a fundamentally new method of extracting 
information about scattering media based on the detection of a fundamental 
property of light, its orbital angular momentum (OAM). Using a generalized Mie 
theory, we have developed a theory for the scattering of light carrying OAM 
from dilute distributions of micro and nanoparticles.

Our analysis is inspired by the fact that optical signals are degraded by 
propagation through scattering media and hence understanding the effect of the 
medium is of critical importance for secure, high-bandwidth communications. We 
show instead that by controlling the axial OAM of the incident beams and 
measuring the OAM of the scattered fields in the far field, the scattering can 
be used to identify the presence of subsets of particles with symmetric or 
chiral distribution within a disordered medium. This may be of particular 
benefit for environmental sensing and metrology, ocean transmissometry, 
nanophotonics, and even applications in biological imaging.

Our method makes use of the fact that structured beams have several 
characteristic length scales and that the component of the total angular 
momentum of light with respect to the direction of propagation of the incident 
beams is, in general, not conserved in order to derive the spatial distribution 
of the scattering particles. This information is obtained maintaining detectors 
and light sources in the same positions, a unique feature that can be extremely 
useful in many applications.
We also found that the signal-to-noise ratio does not degrade as the OAM 
increases, thus confirming that they are an ideal basis for transmitting 
multiplexed signals.

The fundamental nature and the generality of this new theory will open the way 
to new experimental approaches in fields as diverse as nanophtonic and marine 
or atmospheric optics. In particular, this theory could be extended to denser 
distributions by including multiple scattering effects through multi-particle 
Green's functions~\cite{mcarthur17b,mcarthur17c} for applications in nonlinear 
and quantum nanophotonics.

On the other hand, measuring the OAM of the scattered light in the far field 
provides constraints on effective medium models in which the field scattered by 
these distributions of particles is reproduced by equivalent continuous 
dielectric functions, replacing the evaluation of the scattering from many 
particles with a calculation of propagation through an effective medium.

\begin{acknowledgments}
We wish to thank our colleagues D.\ McKee, P.\ Griffin, S.\ Spesyvtseva, P.\ F.\ 
Brevet and H.\ Okamoto for several useful discussions. 
A.M.Y and D.M. thank the Leverhulme Trust for the award of a Leverhulme Trust 
Research Project Grant No. RPG-2017-048. D.M. is supported by the
EPSRC (EP/K503174/1). In agreement with EPSRC policy, datasets will be made 
available on Pure, the University of Strathclyde data repository. A.M.Y. thanks the Natural Environment Research Council for support via award NE/P003265/1.
\end{acknowledgments}

\appendix

\section{\label{app:A1} Multipole expansions}
In this appendix we briefly summarize the relation between spherical harmonics, 
vector spherical waves, multipoles and the eigenfunctions $\mathbf{Y}^m_{jl}$ 
of the orbital angular momentum, spin  and total angular momentum operators
\begin{eqnarray}
  \hat{L} &=& -i \mathbf{r} \times \nabla, \\
  \hat{S} &=& i I \times, \\
  \hat{J} &=& \hat{L}+\hat{S},
\end{eqnarray}
where we have used the representation of the spin operator, valid for photons, 
given in Ref.~\cite{biedenharn85} (Chapter 3, Appendix D) -- with  $I$ the 
identity \added[id=dm]{matrix} in the three dimensional space -- and the 
operators are divided by $\hbar$. 
The  eigenfunctions 
$\mathbf{Y}^m_{jl}$ of $\hat{L}^2,\hat{J}^2,\hat{J}_z$ satisfy
\begin{eqnarray}
  \hat{L}^2\mathbf{Y}^m_{jl} &=& l(l+1)\mathbf{Y}^m_{jl}, \\
  \hat{J}^2\mathbf{Y}^m_{jl} &=& j(j+1)\mathbf{Y}^m_{jl}, \\
  \hat{J}_z\mathbf{Y}^m_{jl} &=& m\mathbf{Y}^m_{jl},
\end{eqnarray}
and can be found in Ref.~\cite{biedenharn85}
\footnote{We note that in Eq.~(6.58) of Ref.~\cite{biedenharn85}, the authors 
use $\hat{\psi}_+=-\mathbf{\hat{e}}_+$ and that there is a factor of $i$ 
between their definition of $\mathbf{N}_{jm}$ and ours.}.
As the spin of light is $s=1$, we need to consider the above relations only for 
the three cases $j=l$, and $j=l \pm 1$. For $j=l, l\pm 1$, the eigenfunctions 
are
\begin{eqnarray}
  \mathbf{Y}^m_{jj}(\Omega) &=& \frac{\hat{L}}{\sqrt{j(j+1)}} Y_{j,m}(\Omega) \label{eq:Y_JJ} \\
  &=&  \mathbf{\hat{e}}_- \alpha^-_{jj} Y_{j,m+1}
    +\mathbf{\hat{e}}_+ \alpha^+_{jj}  Y_{j,m-1}
    +\mathbf{\hat{z}} \alpha^z_{jj} Y_{j,m}  \nonumber \\
  \mathbf{Y}^m_{j,j-1}(\Omega) &=& \frac{j \mathbf{\hat{r}} 
    -i( \mathbf{\hat{r}} \times \hat{L})}{\sqrt{j(2j+1)}} Y_{jm}(\Omega) \label{eq:Y_JJ-1} \\ 
  &=& \mathbf{\hat{e}}_- \alpha^-_{j,j-1}  Y_{j-1,m+1}
    +\mathbf{\hat{e}}_+ \alpha^+_{j,j-1} Y_{j-1,m-1} \nonumber \\
  &&  +\mathbf{\hat{z}} \alpha^z_{j,j-1} Y_{j-1,m} \nonumber \\
  \mathbf{Y}^m_{j,j+1}(\Omega) &=& -\frac{(j+1)\mathbf{\hat{r}}
    +i (\mathbf{\hat{r}} \times \hat{L})}{\sqrt{(j+1)(2j+1)}} Y_{jm}(\Omega) \label{eq:Y_JJ+1} \\ 
  &=& \mathbf{\hat{e}}_- \alpha^-_{j,j+1}  Y_{j+1,m+1}
    +\mathbf{\hat{e}}_+ \alpha^+_{j,j+1}  Y_{j+1,m-1} \nonumber \\
   && +\mathbf{\hat{z}} \alpha^z_{j,j+1} Y_{j+1,m} \nonumber
\end{eqnarray}
where
\begin{subequations}
\begin{eqnarray}
  \alpha^+_{jj} &=& \sqrt{\frac{(j+m)(j-m+1)}{2j(j+1)}} \\
  \alpha^-_{jj} &=&  \sqrt{\frac{(j-m)(j+m+1)}{2j(j+1)}} \\
  \alpha^z_{jj} &=& \frac{m}{\sqrt{j(j+1)}}
\end{eqnarray}
\end{subequations}
\begin{subequations}
\begin{eqnarray}
  \alpha^+_{j,j-1} &=& -\sqrt{\frac{(j+m)(j+m-1)}{2j(2j-1)}} \\
  \alpha^-_{j,j-1} &=& \sqrt{\frac{(j-m)(j-m-1)}{2j(2j-1)}} \\
  \alpha^z_{j,j-1} &=& \sqrt{\frac{(j-m)(j+m)}{j(2j-1)}}
\end{eqnarray}
\end{subequations}
\begin{subequations}
\begin{eqnarray}
  \alpha^+_{j,j+1} &=& -\sqrt{\frac{(j-m+2)(j-m+1)}{2(j+1)(2j+3)}} \\
  \alpha^-_{j,j+1} &=& \sqrt{\frac{(j+m+2)(j+m+1)}{2(j+1)(2j+3)}} \\
  \alpha^z_{j,j+1} &=& -\sqrt{\frac{(j-m+1)(j+m+1)}{(j+1)(2j+3)}}
\end{eqnarray}
\end{subequations}
and
\begin{equation}
  Y_{jm}(\Omega) = \sqrt{\frac{2j +1}{4\pi} \frac{(j-m)!}{(j+m)!}} 
    P_j^m(\cos{\theta}) \exp{(im\varphi)},
\end{equation}
are the scalar spherical harmonics with $P_j^m(\cos{\theta})$ the associated 
Legendre function~\cite{jackson99}. 
The vector spherical harmonics can be expressed in terms of these 
eigenfunctions as
\begin{eqnarray}
  \mathbf{m}_{jm}(\Omega) &=& \mathbf{Y}^m_{jj}(\Omega), \label{eq:sph_m}\\
  \mathbf{n}_{jm}(\Omega) &=& \mathbf{\hat{r}} \times \mathbf{m}_{jm}(\Omega) \nonumber \\
   &=& i\Bigg[\sqrt{\frac{j+1}{2j+1}} \mathbf{Y}^m_{j,j-1}(\Omega) \nonumber \\
   && + \sqrt{\frac{j}{2j+1}} \mathbf{Y}^m_{j,j+1}(\Omega)\Bigg]. \label{eq:sph_n} 
\end{eqnarray}
The orthonormality relations over the $4\pi$ solid angle
\begin{eqnarray*}
  \int Y_{jm}^*Y_{j^\prime m^\prime} d\Omega &=& \delta_{jj^\prime}
    \delta_{mm^\prime}, \\
  \int {\mathbf{Y}^m_{j,l}}^*\mathbf{Y}^{m^\prime}_{j^\prime,l^\prime} d \Omega 
    &=& \delta_{jj^\prime}\delta_{ll^\prime}\delta_{mm^\prime}, \\
  \int \mathbf{m}_{jm}^*\mathbf{m}_{j^\prime m^\prime} d \Omega &=& 
    \int \mathbf{n}_{jm}^*\mathbf{n}_{j^\prime m^\prime} d \Omega = 
    \delta_{jj^\prime}\delta_{mm^\prime}, \\
  \int \mathbf{m}_{jm}^*\mathbf{n}_{j^\prime m^\prime} d \Omega &=& 0,
\end{eqnarray*}
are very useful.

The spherical vector waves corresponding to these angular functions are
\begin{eqnarray}
  \mathbf{M}_{jm}(\mathbf{r}) &=& z_j(k|\mathbf{r}|) \mathbf{m}_{jm}(\Omega),\\
  \mathbf{N}_{jm}(\mathbf{r}) &=& k^{-1}\mathbf{\nabla} \times \mathbf{M}_{jm}
  (\mathbf{r}) \nonumber \\
  &=& i\Bigg[\sqrt{\frac{j+1}{2j+1}}z_{j-1}(k|\mathbf{r}|)\mathbf{Y}^m_{j,j-1}
  (\Omega) \nonumber \\
  && -\sqrt{\frac{j}{2j+1}}z_{j+1}(k|\mathbf{r}|)\mathbf{Y}^m_{j,j+1}(\Omega)\Bigg]
  \nonumber \\
  &=& i \frac{z_j(k|\mathbf{r}|)}{k|\mathbf{r}|} \sqrt{j(j+1)}Y_{j,m}(\Omega) 
  \hat{\mathbf{r}} \nonumber \\
  && + \frac{\p_{k|\mathbf{r}|} \left[k|\mathbf{r}| z_j(k|\mathbf{r}|)\right]}
  {k|\mathbf{r}|} \mathbf{n}_{jm}(\Omega),
\end{eqnarray}
where $z_l$ are either spherical Bessel or Hankel functions of the first kind.
The \replaced[id=dm]{recurrence relations}{identities} $z_{j-1}(r) = z_j^\prime(r) + 
(j+1)z_j(r)/r$ and $z_{j+1}(r) = -z_j^\prime(r) + jz_j(r)/r$ can be used to 
verify the last equality.
From the expression above, we can see that $\mathbf{m}_{jm}$ and 
$\mathbf{M}_{jm}$ are eigenfunctions of $\hat{L}^2$, $\hat{J}^2$ and $\hat{J}_z$, 
but $\mathbf{n}_{jm}$ and $\mathbf{N}_{jm}$ are eigenfunctions only of 
$\hat{J}^2$ and $\hat{J}_z$, as shown by the presence of $\mathbf{Y}^m_{j,j-1}$ 
and $\mathbf{Y}^m_{j,j+1}$. This happens because the vector product and the rotor 
mix states with different orbital angular momentu\replaced[id=dm]{a}{m} $l$.

Finally, the electric and magnetic multipoles are exact solutions of Maxwell's 
equations used in the expansion of plane waves. They are defined as
\begin{eqnarray}
  \mathcal{J}^H_{jm}(\mathbf{r}) &=& \left[ 
    \begin{array}{c} 
      \mathbf{M}_{jm} (\mathbf{r}) \\ 
      -i \sqrt{\frac{\varepsilon_r}{\mu_r}} \mathbf{N}_{jm} (\mathbf{r})
    \end{array} \right], \label{eq:sph_multi_H} \\
  \mathcal{J}^E_{jm}(\mathbf{r}) &=& \left[
    \begin{array}{c} 
      \mathbf{N}_{jm}(\mathbf{r})  \\  -i
      \sqrt{\frac{\varepsilon_r}{\mu_r}} \mathbf{M}_{jm}(\mathbf{r})
    \end{array} \right] \label{eq:sph_multi_E}, 
\end{eqnarray}
with the Bessel functions used for $\mathbf{M}_{jm}$, $\mathbf{N}_{jm}$. The 
electric and magnetic multipoles, $\mathcal{S}^H_{jm}$ and $\mathcal{S}^E_{jm}$, 
used in the expansion of scattering fields are defined analogously, but using 
the Hankel functions instead of the Bessel functions.

\section{\label{app:A2} Plane wave expansion coefficients}
The beam expansion coefficients of the plane wave in Eqs.~(\ref{eq:pw_bec_h}-
\ref{eq:pw_bec_e}) are
\begin{eqnarray}
  a^H_{jm} &=& 4\pi i^j \mathbf{m}^*_{jm}(\Omega_0) \cdot \mathbf{E}^0, \\
  a^E_{jm} &=& \sqrt{\frac{\mu_r}{\varepsilon_r}}
    4\pi i^{j+1} \mathbf{m}^*_{jm}(\Omega_0) \cdot \mathbf{H}^0, 
\end{eqnarray}
with $\Omega_0=(\theta_0, \varphi_0)$ the angular spherical coordinates of wave 
vector $\mathbf{k}_0$ and $\mathbf{m}_{jm}(\Omega_0)$ is the vector spherical 
harmonic defined as in Jackson~\cite{jackson99}. For the special case in which 
$\mathbf{k}_0$ is parallel to the $z$ axis, we have $\Omega_0=(0,\cdot)$ and
the only non-vanishing spherical harmonics are
\begin{equation}
  Y_{j,0}(0,\cdot) = \sqrt{\frac{2j+1}{4 \pi}},
\end{equation}
which, together with Eqs.~(\ref{eq:Y_JJ}-\ref{eq:sph_m}), means that the 
longitudinal waves with electric or magnetic fields parallel to the propagation
axis $z$ do not induce scattering. For transverse waves propagating along $z$, 
$m$ has values $\pm 1$, depending on the polarization, and
\begin{eqnarray}
  a^H_{j,\mp 1} &=& i^j \sqrt{4 \pi(2j+1)}\frac{E^0_x \pm i E^0_y}{2} \nonumber \\
    &=& i^j \sqrt{4 \pi(2j+1)}\frac{\mathbf{\hat{e}}_{\pm} \cdot 
      \mathbf{E}^0}{\sqrt{2}}, \label{eq:pw_ah} \\
  a^E_{j,\mp 1} &=& \sqrt{\frac{\mu_r}{\varepsilon_r}}i^{j+1} 
    \sqrt{4 \pi(2j+1)} \frac{H^0_x \pm i H^0_y}{2} \nonumber \\ 
    &=& \sqrt{\frac{\mu_r}{\varepsilon_r}}i^{j+1} \sqrt{4 \pi(2j+1)} 
    \frac{\mathbf{\hat{e}}_{\pm} \cdot \mathbf{H}^0}{\sqrt{2}}, \label{eq:pw_ae}
\end{eqnarray}
with $\mathbf{\hat{e}}_{\pm} = (\mathbf{\hat{x}} \pm i \mathbf{\hat{y}})/\sqrt{2}$.

For an incident field with the angular spectrum representation 
\begin{eqnarray}
  \mathbf{E}(\mathbf{r}) &=& \int \tilde{\mathbf{E}}(\Omega_k) 
    \exp{\left[i \mathbf{k}(\Omega_k) \cdot \mathbf{r}\right]} d \Omega_k, \\
  \mathbf{H}(\mathbf{r}) &=& \int \tilde{\mathbf{H}}(\Omega_k)
    \exp{\left[i \mathbf{k}(\Omega_k) \cdot \mathbf{r}\right]} d \Omega_k, 
\end{eqnarray}
with $\tilde{\mathbf{H}} = \sqrt{\varepsilon_r/\mu_r}(\mathbf{\hat{k}} \times \tilde{
\mathbf{E}})$, the beam expansion coefficients for a particle centered in 
$\mathbf{r}$ are obtained by considering the angular spectrum representation in 
the reference frame centered in $\mathbf{r}_t$ and using the beam expansion 
coefficients of the plane waves,
\begin{eqnarray}
  A^H_{\nu}(\mathbf{r}_t) &=& 4\pi i^j \int \mathbf{m}^*_{\nu}(\Omega_k)
    \cdot \tilde{\mathbf{E}}(\Omega_k) \nonumber \\
    && \times \exp{[-i \mathbf{k}(\Omega_k)\cdot \mathbf{r}_t]} d \Omega_k, \\
  A^E_{\nu}(\mathbf{r}_t) &=& \sqrt{\frac{\mu_r}{\varepsilon_r}} 4\pi i^{j+1}
    \int \mathbf{m}^*_{\nu}(\Omega_k) \cdot \tilde{\mathbf{H}}(\Omega_k) \nonumber \\
    && \times\exp{[-i \mathbf{k}(\Omega_k) \cdot \mathbf{r}_t]} d \Omega_k.
\end{eqnarray}
%

\section{\label{app:A3} Asymptotic expressions}
We use the far field asymptotic expansions 
\begin{eqnarray}
  \mathbf{M}_{jm}(\mathbf{r}') &\sim& \frac{\exp{(ik|\mathbf{r}'|)}}{|\mathbf{r}'|} 
    (-i)^{j+1} \mathbf{m}_{jm}(\theta',\varphi') + \mathcal{O}\left(\frac{1}{r^2}\right) \nonumber \\
    &\sim& \frac{\exp{(ik|\mathbf{r}|)}}{|\mathbf{r}|}
    \exp{(-ik \hat{\mathbf{r}} \cdot \mathbf{r}_t)}  
    (-i)^{j+1} \mathbf{m}_{jm}(\theta,\varphi) \nonumber \\
    &&+ \mathcal{O}\left(\frac{1}{r^2}\right) \label{eq:A3-1} \\
  \mathbf{N}_{jm}(\mathbf{r}') &\sim& \frac{\exp{(ik|\mathbf{r}'|)}}{|\mathbf{r}'|} 
    (-i)^{j} \mathbf{n}_{jm}(\theta',\varphi') + \mathcal{O}\left(\frac{1}{r^2}\right) \nonumber \\
    &\sim& \frac{\exp{(ik|\mathbf{r}|)}}{|\mathbf{r}|} 
    \exp{(-ik \hat{\mathbf{r}} \cdot \mathbf{r}_t)} 
    (-i)^{j} \mathbf{n}_{jm}(\theta,\varphi)\nonumber \\
    && + \mathcal{O}\left(\frac{1}{r^2}\right), \label{eq:A3-2}
\end{eqnarray}
where $\mathbf{r}'=\mathbf{r}-\mathbf{r}_t$, $\hat{\mathbf{r}} = \mathbf{r}/
|\mathbf{r}|$ and $|\mathbf{r}| \rightarrow \infty$.
\replaced[id=dm]{Employing the Jacobi-Anger identity gives the following expansion}
{From the Jacobi-Anger identity is}
\begin{eqnarray}
  \exp(-ik \hat{\mathbf{r}} \cdot \mathbf{r}_t) &=& \exp\{-ik |\mathbf{r}_t|
    [\cos{\theta_t}\cos{\theta} \nonumber \\ 
  && + \sin{\theta_t}\sin{\theta} \cos{(\varphi - \varphi_t)]}\} \nonumber \\
  &=& \exp{(-ik |\mathbf{r}_t| \cos{\theta_t} \cos{\theta} )} \nonumber \\
  && \times \sum_{n=-\infty}^{\infty} (-i)^n J_n( k |\mathbf{r}_t| 
    \sin{\theta_t} \sin{\theta}) \nonumber \\
  && \times \exp{[i n (\varphi - \varphi_t)]}. \label{eq:transl}
\end{eqnarray} 
The asymptotic form of the spherical multipoles is 
\begin{eqnarray}
  \mathcal{S}^{H \infty}_{jm}(\mathbf{r}) &=& \frac{\exp{(ik |\mathbf{r}|)}}{|\mathbf{r}|}
    \left[
    \begin{array}{c} 
      (-i)^{j+1} \mathbf{m}_{jm}(\theta,\varphi)  \\
      -i \sqrt{\frac{\varepsilon_r}{\mu_r}} 
      (-i)^{j} \mathbf{n}_{jm}(\theta,\varphi) 
    \end{array} \right] \nonumber \\
  &=& \frac{\exp{(ik |\mathbf{r}|)}}{|\mathbf{r}|}
    \mathcal{P}_{jm}^H(\theta, \varphi) , \label{eq:svwf-ff-E} \\
  \mathcal{S}^{E \infty}_{jm}(\mathbf{r}) &=&  
    \frac{\exp{(ik |\mathbf{r}|)}}{|\mathbf{r}|} \left[ 
    \begin{array}{c} 
      (-i)^{j} \mathbf{n}_{jm}(\theta,\varphi) \\  
      -i \sqrt{\frac{\varepsilon_r}{\mu_r}}(-i)^{j+1} \mathbf{m}_{jm}(\theta,\varphi)
    \end{array} \right] \nonumber \\
   &=& \frac{\exp{(ik |\mathbf{r}|)}}{|\mathbf{r}|} \mathcal{P}_{jm}^E(\theta,\varphi) \label{eq:svwf-ff-H}
\end{eqnarray}
where $\mathcal{P}$ contain the angular dependence of the $\mathbf{m}$ and 
$\mathbf{n}$ harmonics. The dependence upon $\varphi$ in Cartesian coordinates 
can be found using Eqs.~(\ref{eq:Y_JJ}-\ref{eq:Y_JJ+1}).
For $m=\pm1$ and $\theta << 1$, to the fist order in $\theta$, we have 
\begin{eqnarray}
  \mathbf{m}_{j,\pm 1}(\theta,\varphi) &\sim& 
    \mathbf{\hat{e}}_{\pm1} \frac{ Y_{j,0}}{\sqrt{2}} \pm
    \mathbf{\hat{z}} \frac{Y_{j,\pm 1}}{\sqrt{j(j+1)}}, \label{eq:small_theta_m} \\ 
  \mathbf{n}_{j,\pm 1}(\theta,\varphi) &\sim& \mp \frac{i Y_{j,0} }{2}
    \big[\mathbf{\hat{e}}_{\pm 1} \sqrt{2} \cos{\theta} \nonumber \\
    &&-\mathbf{\hat{z}} \sin{\theta} \exp{(\pm i \varphi)}\big]. \label{eq:small_theta_n}
\end{eqnarray}


%

\end{document}